\documentclass[prl,
 reprint,
 amsmath,amssymb,
 aps,
]{revtex4-2}

\usepackage{graphicx}
\usepackage{dcolumn}
\usepackage{bm}

\usepackage[utf8]{inputenc}
\usepackage[english]{babel}
\usepackage[dvipsnames]{xcolor}
\usepackage{hyperref}

\usepackage{amsmath,amssymb}
\usepackage{mathtools}
\usepackage{mathrsfs}
\usepackage{physics}
\usepackage{dsfont}
\usepackage{svg}
\usepackage[percent]{overpic} 

\usepackage{soul}

\usepackage{tikz}
\usetikzlibrary{arrows.meta, positioning}
\usetikzlibrary{quantikz2}

\usepackage{fancyhdr}

\usepackage[caption=false]{subfig} 

\ExplSyntaxOn
\makeatletter
\newsavebox{\@bra}
\newsavebox{\@brb}
\DeclarePairedDelimiterX\myinnerp[1]{.}{.}{
  \delimsize\langle
  \hspace*{0.3mm}\hspace*{0.55mm}
  \savebox{\@bra}{$\left\langle\vphantom{#1}\right.$}
  \hspace*{-0.9\wd\@bra}
  \delimsize\langle
  #1
  \delimsize\rangle
  \hspace*{0.3mm}\hspace*{0.55mm}
  \savebox{\@brb}{$\left.\vphantom{#1}\right\rangle$}
  \hspace*{-0.9\wd\@brb}
  \delimsize\rangle
}
\makeatother
\ExplSyntaxOff

\begin{document}


\title{Finite steps optimise dissipation in stochastically controlled quantum systems}

\author{Theodore McKeever}
\author{Harry J. D. Miller}
\email[Corresponding author: ]{harry.miller@manchester.ac.uk}
\author{Ahsan Nazir}

\affiliation{Department of Physics and Astronomy, The University of Manchester, Manchester M13 9PL, UK.}
\date{\today}

\begin{abstract}
Motivated by the need for precise, energy-efficient, and experimentally realistic quantum control protocols, we investigate the thermodynamic cost of performing quantum step-equilibration processes under the influence of classical stochastic control fields. Whereas purely deterministic protocols exhibit dissipation that scales inversely with the number of steps, we show that weak Gaussian noise in the control variables induces dissipative contributions that grow linearly with the number of steps. Consequently, we derive the finite optimal number of steps and minimal achievable average dissipated work and its variance using the quantum thermodynamic length. These results are demonstrated using two paradigmatic examples: a Landau-Zener sweep of a qubit strongly coupled to a thermal bath and the erasure of a transverse-field Ising model.
\end{abstract}

\maketitle



\textit{Introduction}---Understanding the energetic cost of controlling quantum systems remains a central problem in nonequilibrium thermodynamics. Extensive work has been conducted in establishing links between dissipation and control through fluctuation relations and entropy production bounds, both in classical stochastic \cite{Seifert_2012}  and quantum thermodynamics \cite{ binder2018thermodynamics,campbell2026roadmap}. A central problem in thermodynamic optimal control is to find protocols that minimise dissipation, which has led to a variety of optimisation approaches and explorations of thermodynamic trade-offs in the quantum regime \cite{cavina2018optimal,abiuso2020geometric,abiuso2020optimal,brandner2020thermodynamic,erdman2023pareto,van2021geometrical,van2023thermodynamic,zulkowski2015geometry}.







Optimal protocols in quantum thermodynamics are typically formulated under the assumption of perfect control. However, realistic control fields are never perfectly stable: classical noise arising from fluctuating fields, finite electronic bandwidths, or feedback imperfections \cite{clerk2010introduction, Wiseman_Milburn_2009} are ubiquitous in experimental platforms. This includes superconducting qubits \cite{koch2007charge, krantz2019quantum}, trapped ions \cite{barreiro2011open, home2009complete}, quantum dots \cite{koong2025coherentcontrolquantumdotspins} and cold atoms \cite{myatt2000decoherence}. Therefore, quantifying how such imperfections translate into thermodynamic costs is essential for both designing thermodynamically efficient protocols and benchmarking the practical limits of quantum control fidelity across a wide range of experiments and technologies.

The thermodynamic cost of stochastic control has been considered in classical scenarios \cite{Guéry-Odelin_2023,ChenStochastic2020,Nerirandomtime,MachtaDissipationBound,Large_Chetrite_Sivak_2018}. Key results~\cite{MachtaDissipationBound,Large_Chetrite_Sivak_2018,BryantAutonomous} from adiabatic linear response theory have derived fundamental control costs based on thermodynamic length \cite{PhysRevLett.99.100602,sivak2012thermodynamic}, establishing a non-vanishing contribution to entropy production that persists even in the quasi-static limit. For quantum systems, much less is known, largely due to the difficulty of consistently accounting for the interplay between noise, control, and environmental timescales when modelling classical control noise in driven open quantum systems. Here, we develop a stochastic linear response framework for quantum step-equilibration processes subject to  weak Gaussian noise. Practically, step-equilibration processes can be implemented  in many experimental platforms \cite{MeinertQuenchColdAtoms,GuardadoColdAtoms,Jurcevic2014TrappedIons,Guo2019SuperconductingQ}, and serve as a paradigmatic model for quantum thermodynamics \cite{Anders_2013,Gallego_2014,Baumer2019imperfect,MillerMehboudi2020}. By discretising protocols into separate quench and thermalisation steps, we derive  expressions for both deterministic and stochastic components of the average  dissipated work, revealing a finite-time optimum step number in which total dissipation is minimised. This optimal step number is determined geometrically by the quantum thermodynamic length, thus establishing a new relationship between quantum geometry and stochastic control. In addition, we  derive an optimal step number for minimising the work fluctuations obtained from a two-point measurement \cite{esposito2009nonequilibrium} of the process. This step number differs from the optimal solution for average dissipation due to the presence of non-commutativity and quantum friction \cite{quantumWorkStats,PhysRevLett.123.230603}, indicating a richer control landscape for quantum systems in comparison to classical stochastic control.



\begin{figure}
        \centering
        \includegraphics[clip, trim=2cm 4.4cm 0.5cm 3.6cm,width=0.98\linewidth]{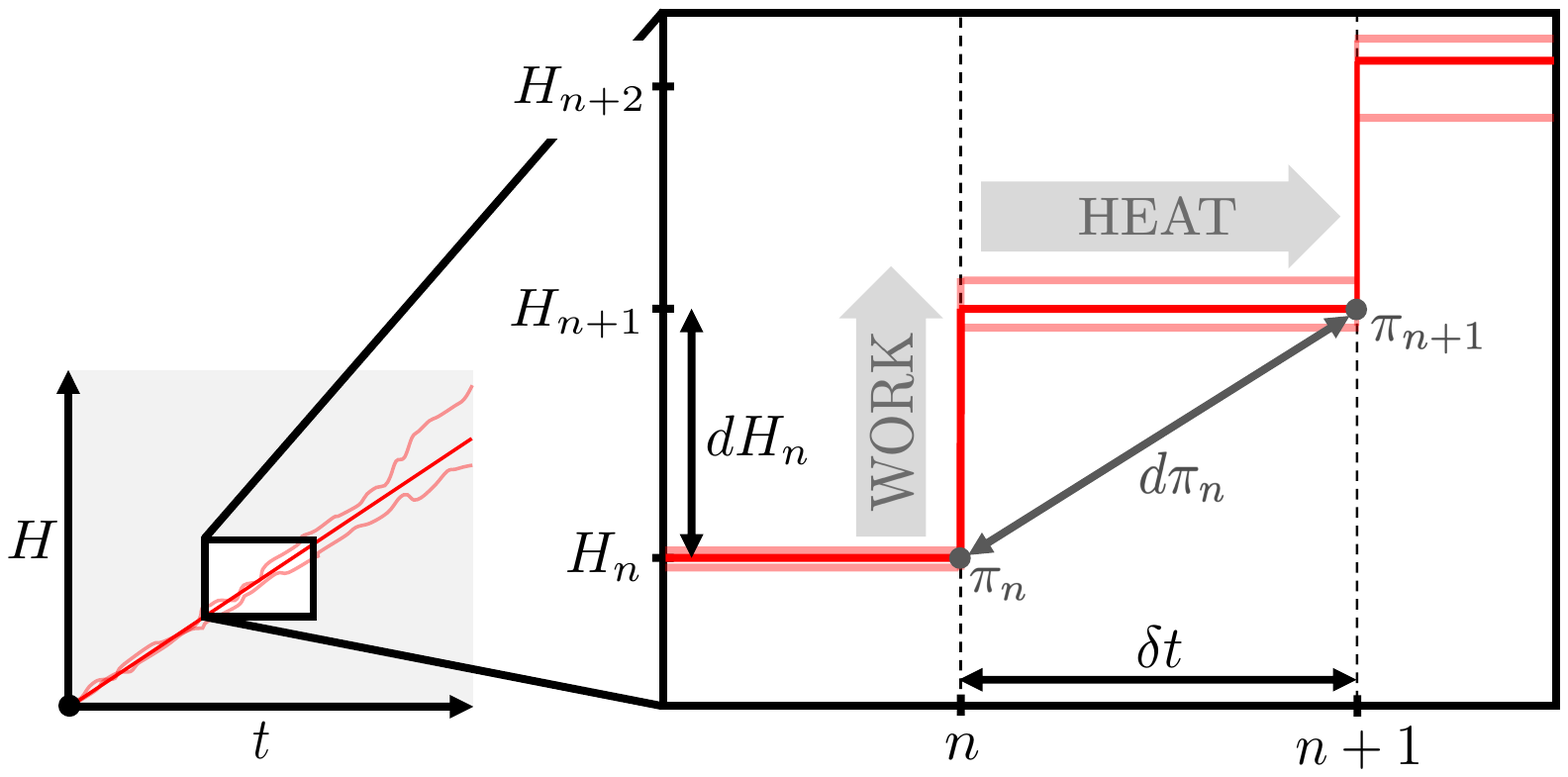}
    
        \caption{Illustration of the quench process with each quench step having two stages, sudden perturbation and then thermalisation over characteristic timescale $\delta t$, where work and heat are exchanged respectively. The controls follow the red line, while faint red lines indicate alternative stochastic trajectories. Double-arrowed lines denote small changes. 
        \label{fig:quenchmodel}}
\end{figure}

\textit{Step-equilibration processes}---We begin by considering a controlled quantum system in contact with a thermal bath at inverse temperature $\beta$ undergoing discrete quench protocols in the absence of control noise where the initial and final target Hamiltonians $H_0$ and $H_N$ are fixed and quenches are applied incrementally over $N$ time steps \cite{nulton1985quasistatic,PhysRevLett.99.100602,Anders_2013,Gallego_2014,Baumer2019imperfect,quantumWorkStats}. The number of steps $N$ functions as a direct measure of protocol duration and at each $0\leq n \leq N $ the system is given adequate time $\delta t$ to thermalise (Fig.~\ref{fig:quenchmodel}). This represents a timescale separation between control and equilibration.
This model is sufficiently general to  accommodate strong system-bath coupling and large many-body systems, and approximates a continuous-time process in the limit $N\gg 1$. 

At the beginning of the $n\textsuperscript{th}$ quench step, the state is assumed to be in a Gibbs state with respect to $H_n$, denoted by the density operator $\pi_n = e^{-\beta H_n}/Z_n$ where $Z_n=\Tr[e^{-\beta H_n}]$ is the partition function with respect to $H_n$. Since each quench is unitary, the mean work done on the system is taken to be the total energy change $W_n = \Tr [d H_{n}  \pi_n]$ according to the change $d H_n = H_{n+1}-H_n$. The relevant thermodynamic cost of this step-equilibration process is quantified by the dissipated work $W_\text{diss}=W-\Delta F$, where $\Delta F=-\beta^{-1} \text{log} (Z_N/Z_0)$ is the change in equilibrium free energy. One can show that the total dissipated work after the $N$\textsuperscript{th} quench is completed  is expressible in terms of the summed quantum relative entropy $S(A \| B)= \Tr [ A(\log A - \log B) ]$ between successive thermal states \cite{parrondo2015thermodynamics,scandi2019thermodynamic,quantumWorkStats}, 
\begin{align}\label{eq:relent}
    W_\text{diss}& = \frac{1}{\beta}\sum_{n=0}^{N-1} S(\pi_{n}\| \pi_{n+1})\geq 0, 
\end{align}
with positivity reflecting the second law of thermodynamics. 

As an optimal control problem we aim to determine the sequence of perturbative quenches between $H_0$ and $H_N$ within a given number of steps $N$ that minimises $W_\text{diss}$. In such regimes one may apply adiabatic linear response theory which allows application of thermodynamic geometry to solve this optimisation. In the case of deterministic control, a tight lower bound on dissipation is given by
\begin{align}\label{eq:length}
    W_\text{diss}\geq \frac{\mathcal{L}^2}{2N} ,
\end{align}
where the thermodynamic length $\mathcal{L}$ is a geometric distance on the space of Gibbs states connecting the initial and final configurations, and will be defined later. 
The key point is that the bound is saturated by choosing a deterministic protocol that maintains equal thermodynamic length along each step, scaling \textit{inversely} with the number of steps. Taking $N\to \infty$ we obtain vanishing dissipation in accordance with the reversible quasi-static limit. In the next section, we prove that this inverse scaling breaks down in the presence of stochastic control.

\textit{Stochastic driving}---We now extend the above deterministic framework to include noise on the controlled quenches. An analytical average over the resulting sample paths then yields an expression for $\myinnerp*{W_\text{diss}}$, where $\myinnerp*{ ... }$ denotes a noise-average. 
A general noise process is introduced into the model, following the decomposition in \cite{budini2001quantum}, where the system Hamiltonian is comprised of deterministic and stochastic parts acting on a set of $d$ operators $\{V_j\}_{j=1}^d$, such that 
\begin{align}\label{eq:Ham}
    H_n & \mapsto H_n + \Tilde{H}_n =  V_j (v_n^j + \xi_n^j)
\end{align} 
where Einstein summation notation is used. The stochastic evolution of the system Hamiltonian is contained within the noise increments $d\Tilde{H}_n$ and the structure of the $v_n^j$, $\xi_n^j$ and $V_j$ preserves the Hermiticity. Without loss of generality, we take all $V_j$ to be Hermitian, therefore the $v_n^j$ represent real-valued deterministic magnitudes, and the $\xi_n^j$ real-valued noise processes.

We are here adopting a sample-and-hold  model in which the noise acts instantaneously with $d\Tilde{H}_n$ during the sudden perturbation stage of each quench, and then is held constant throughout the subsequent equilibration procedure, so the system thermalises with respect to $H_n + \Tilde{H}_n$. Conceptually, the leading contribution to the noise can be interpreted as arising through deviations from the intended parameter value at each quench, with the bath timescales much shorter than that of control noise.  In order to mirror the deterministic scenario outlined previously, we will treat each noisy quench as a perturbation relative to the thermal energy scales, so that heuristically $\beta \norm*{ dH_n+d\tilde{H}_n} \ll 1~ \forall n$ for a suitably chosen norm. We also make a further weak noise approximation, $\norm*{\Tilde{H}_n} \ll \norm*{H_n} ~\forall n$; this regime is reasonable for many current platforms that admit a high degree of control. One may then apply adiabatic linear response theory to \textit{small} quantities: $\{\norm*{dH_n}, \norm*{d\Tilde{H}_n} , \norm*{\Tilde{H}_n} \}\in \mathcal{O}(\epsilon)$ with $\epsilon$ the relevant energy scale. We derive the  expansion up to second order in Appendix~\ref{app:noise}, and find
\begin{align}\label{eq:WdissGeo}
    \myinnerp*{W_ \text{diss}}  = \frac{1}{2}\sum_{n=0}^{N-1}  ~  g_{ij}^{(n)}\bigg[ dv^i_n dv^j_n + \myinnerp*{ \Delta\xi_n^i \Delta\xi_n^j} \bigg]+\mathcal{O}(\epsilon^3)
\end{align}
where $\Delta\xi_n^i=\xi_{n+1}^i-\xi_{n}^i$ is the stochastic noise increment associated with the $n$\textsuperscript{th} quench of control parameter $i$. Here $g^{(n)}_{ij}$ is the $d$-dimensional thermodynamic metric on the space of deterministic control variables $\vec{v}_n=\{v_n^j\}_{j=1,2,\cdots,d}$ \cite{scandi2019thermodynamic}, defined in terms of the Kubo covariance \cite{quantumWorkStats}
\begin{align}\label{eq:KuboMetric}
     g^{(n)}_{ij}&=\beta\int^1_0 ds \ \Tr\left[V_i \pi_n^s V_j \pi_n^{1-s}\right]-\Tr\left[V_i\pi_n\right]\Tr\left[V_j\pi_n\right].
\end{align}
The leading contribution due to the addition of noise comes from the correlation matrix $\myinnerp*{ \Delta\xi_n^i \Delta\xi_n^j}$. 

We now turn our focus to studying the quasi-static limit in which the number of steps approaches $N \gg 1$. To do so we introduce the  smooth curve in the parameter space $\vec{v}_n\mapsto \vec{v}_t$ with parameter $t\in[0,1]$ such that $\vec{v}_{t=0}=\vec{v}_0$ and $\vec{v}_{t=1}=\vec{v}_N$. Derivatives in control variables are replaced according to $d\vec{v}_n \mapsto \Dot{\vec{v}}_t /N$, and the metric tensor becomes  $g_{ij}^{(n)}\mapsto g_{ij}(\vec{v}_t) $. Lastly we define the interpolated variance in noise increments by the tensor
\begin{align}
    \Phi_t:=\Big[\myinnerp*{ \Delta\xi_t^i \Delta\xi_t^j}\Big]_{i,j=1}^d\geq 0,
\end{align}
which is interpolated such that $\Phi_t\simeq \myinnerp*{ \Delta\xi_n^i \Delta\xi_n^j} $ for $t\in[n/N, (n+1)/N]$. In general one can consider a variety of weak noise models including non-Markovian ones, which we review in Appendix~\ref{app:markov}. 

For $N\gg 1$, replacing the discrete sum with a Riemann integral over $t$, Eq.~\eqref{eq:WdissGeo} becomes
\begin{align}\label{eq:WdissCont}
    \myinnerp*{W_ \text{diss}}  \simeq \frac{1}{2}\int^1_0 dt \  \left( \frac{1}{N}  g_{ij}(\vec{v}_t) \Dot{v}_t^i\Dot{v}_t^j + N  g_{ij}(\vec{v}_t)\Phi^{ij}_{t} \right) .
\end{align}
Whilst $W_\text{diss}$ scales with 1/$N$ in the absence of classical noise in accordance with the quasi-static limit, stochastic control yields a positive term linear in $N$. Thus, the quasi-static limit now gives diverging dissipation in the presence of control noise.

The divergence in the quasi-static limit implies that there must be an optimal number of steps $N_\text{opt}$ that minimises the dissipation. This can be found via the following geometric argument. First, we define the  thermodynamic length along the deterministic control path $\Lambda:t\mapsto \vec{v}_t$,
\begin{align}\label{eq:thermolength}
    \mathcal{L}(\Lambda) = \int_0^{1} dt \sqrt{ g_{ij}(\vec{v}_t) \Dot{v}_t^i  \Dot{v}_t^j }  ,
\end{align}
in keeping with Eq.~\eqref{eq:length}. Then, we can apply the Cauchy-Schwarz inequality, $|\langle a,b \rangle|^2 \leq \langle a,a \rangle^2 \langle b,b \rangle^2$, to derive a lower bound on $\myinnerp*{W_ \text{diss}}$, leading to 
\begin{align}\label{eq:WdissBound}
    \myinnerp*{W_ \text{diss}}  \geq \frac{\mathcal{L}(\Lambda)^2}{2N} ~+~ \frac{N}{2} \int^1_0 dt  ~ g_{ij}(\vec{v}_t)\Phi^{ij}_{t} .
\end{align}
If $\Phi_{t}$ is independent of the deterministic parameter velocity, this bound is saturated for any fixed $N$ by a protocol satisfying the geodesic equation. We label the solution by the curve $\Lambda^*:t\mapsto v^*_t$.   
An optimum duration for a specified protocol can be predicted straightforwardly by calculating the point at which the derivative of $\myinnerp*{W_ \text{diss}}$ with respect to $N$ is zero. For constant $\Phi_{t}$, this corresponds to the point where the deterministic and stochastic contributions in Eq.~\eqref{eq:WdissBound} are equal. The optimal number of steps is given by
\begin{align}\label{eq:Nopt}
    N_\text{opt}= \frac{\mathcal{L}(\Lambda^*)}{\sqrt{ 
    \int^1_0  dt \   g_{ij}(\vec{v}_t^{*}) \Phi^{ij}_{t} ~}}.
\end{align}
Inserting this optimum duration back into Eq.~\eqref{eq:WdissBound} reveals the minimum dissipation cost $\myinnerp*{W_ \text{diss}} \geq W_\text{opt}$,
\begin{align}\label{eq:Wopt}
    W_\text{opt} =  \mathcal{L}(\Lambda^*)\sqrt{  \int^1_0 dt  \  \Phi^{ij}_{t} g_{ij}(\vec{v}_t^*)} .
\end{align}
Equations~\eqref{eq:Nopt} and~\eqref{eq:Wopt} are the main results of this paper; for non-vanishing noise, there exists a finite number of steps $N_\text{opt}<\infty$ that achieves the minimum cost $W_\text{opt}>0$. This gives the stochastic refinement of the deterministic geometric bound~\eqref{eq:length}, and can be viewed as a quantum generalisation of classical bounds derived in \cite{MachtaDissipationBound,Large_Chetrite_Sivak_2018}.

\begin{figure}
        \centering
        \begin{overpic}[clip, trim=0cm 0.1cm 0cm 0cm, width=\columnwidth]{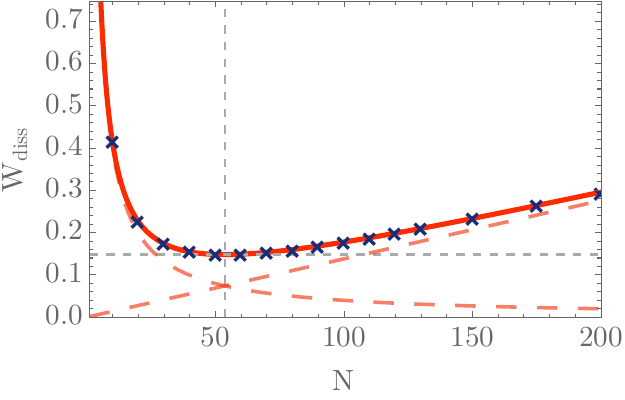}
        \put(44,29.5){\includegraphics[width=0.47\columnwidth]{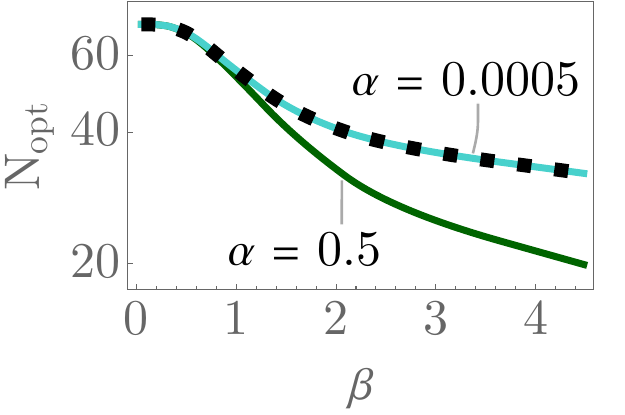}
        }
        \end{overpic}
        \caption{Dissipated work of a qubit avoided crossing against total number of quench steps, $N$, for a GWN process. The red lines represent analytic calculation of $W_\text{diss}$ and its two components (dashed). The dark blue data points display results from numerical trajectory simulations and faint gray dashed lines indicate predictions of $N_\text{opt}$ and $W_\text{opt}$ in units of $[\Delta]$. Inset: $N_\text{opt}$ increases with temperature, plateauing either side of $\beta=1$. The black dotted line coinciding with the turquoise line ($\alpha = 0.0005$) represents the weak-coupling assumption. Parameters: $\beta=1$, $r = 50$, $\sigma_\eta = 0.05$, $\alpha =0.5$, $\Delta =1$, $\omega_0=-5$, $\omega_1=5$.} 
        \label{fig:qubit}
\end{figure}

\textit{Qubit}---To corroborate the predictions \eqref{eq:WdissCont}, \eqref{eq:Nopt} and \eqref{eq:Wopt}, we examine a Landau-Zener ramping on a two-level system with intrinsic tunnelling and time-dependent control of the energy splitting, $H_t = (\omega_t \sigma_z + \Delta \sigma_x )/2$ where $\sigma_i$ represent the Pauli matrices. To illustrate the generality of our approach, we choose to look at the strong system-reservoir coupling regime following the polaron method \cite{NazirCorrelation,JangCoherent,McCutcheonConsistent}. The boundary between system and bath is redrawn by dressing the system states with vibrational modes giving rise to the polaron system Hamiltonian $H_t\mapsto H_t^P = (\omega_t \sigma_z + \Delta \gamma (\beta) \sigma_x )/2$ \cite{strongWorkStats}. The renormalisation constant $\gamma(\beta) := \exp{ -2 \int_0^\infty d\omega J(\omega) \coth{(\beta \omega/2)}/\omega^2}$ suppresses tunnelling, with its effect more severe at higher temperatures and stronger system-bath couplings. Dissipated work is fully captured by changes in the polaron Hamiltonian $H_t^P$ and the system equilibrium state is given by the mean-force Gibbs state \cite{SubasiEquilibrium,TrushechkinOpen,Cresser2021} with respect to $H_t^P$. We adopt a super-Ohmic spectral density $J(\omega) = \alpha \exp (-\omega/\omega_C) \omega^3/\omega_C^2$ with cutoff frequency $\omega_C$ and system-bath interaction strength $\alpha$, which is suitable to describe solid-state systems coupled to acoustic phonons \cite{Nazir_2016}. With boundary conditions set by $\ \omega_0 $ and $ \omega_1 $, alongside the relevant metric tensor derived from Eq.~\eqref{eq:WdissGeo}, the geodesic path can be solved.

Here we focus on Gaussian white noise (GWN) processes \cite{van2007stochastic}. 
Numerically, these processes are constructed using GWN increments where $\xi_n$ are i.i.d. samples with $\xi_n\sim  \mathcal{N}(0,\sigma^2_\eta)$.
From this,
we can express the variance of noise increments as $\Phi_n^\text{GWN } = \myinnerp*{ \Delta\xi_n \Delta\xi_n } = 2\sigma_\eta^2$. 
To assess the legitimacy of our analytic treatment, we numerically simulate $r$ trajectories of this noise process, compute the exact dissipation using Eq.~\eqref{eq:relent} and then examine the average statistics.

Results are shown in Fig.~\ref{fig:qubit}, which displays agreement between numerical trajectory simulations and the linear response predictions \eqref{eq:WdissCont}, \eqref{eq:Nopt} and \eqref{eq:Wopt}, including the emergence of the finite $N_\text{opt}$. The limits of our linear response approximations have been probed for this setup, where temperatures of at least $\beta = 1$ are reliably described, and noise at least as strong as $\Phi_t = 0.01$. The inset of Fig.~\ref{fig:qubit} shows that $N_\text{opt}$ increases with temperature and becomes constant at extreme temperatures. Outside of the high temperature limit $\beta \ll 1$, $N_\text{opt}$ deviates for different system-bath coupling strengths, with stronger coupling favouring shorter protocols. This behaviour underscores the importance of explicitly accounting for strong-coupling effects when optimising quantum control. Although increasing $\alpha$ lowers the optimal number of steps, this does not imply any thermodynamic advantage to stronger system–bath coupling: for larger $\alpha$, $W_\text{diss}$ is higher for all $N$.

\begin{figure}[ht]
    \centering
    \begin{overpic}[clip, trim=0cm 0.5cm 0cm 0cm, width=\columnwidth]{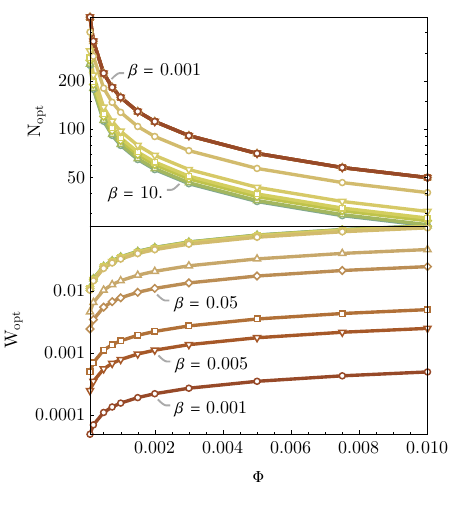}
        \put(45,67){\includegraphics[width=0.44\columnwidth]{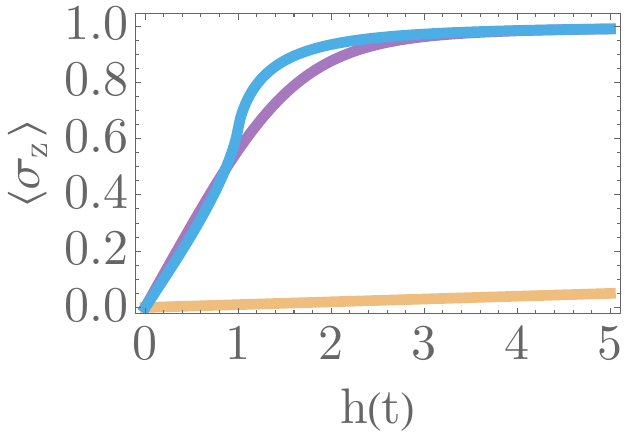}}
    \end{overpic}
     \caption{Optimal step number $N_\text{opt}$ and corresponding minimum dissipated work $W_\text{opt}$ for a TFIM erasure protocol as functions of the noise variance $\Phi$. The optimal work scales as $W_\text{opt} \propto \sqrt{\Phi}$, in agreement with Eq.~\eqref{eq:Wopt}, while $N_\text{opt}$ decreases monotonically with increasing $\Phi$, as predicted by Eq.~\eqref{eq:Nopt}. Increasing temperature leads to larger $N_\text{opt}$ and reduced $W_\text{opt}$. Results are for $L=180$. 
    Inset: Magnetisation per spin as a function of $h$ ($h_1=5.0$) for $\beta=0.01$ (orange), $\beta=1.0$ (purple), and $\beta=100$ (blue), in the thermodynamic limit $L \to \infty$.}
    \label{fig:isingVsPhi}
\end{figure}

\textit{Transverse-field Ising model (TFIM)}---The TFIM provides another paradigmatic setting to explore a step-equilibration process. It is constructed as a chain of $L$ binary spin degrees of freedom subjected to a transverse field of strength $h= h_n + \Tilde{h}_n$ \cite{pfeuty1970one,Sachdev_2011} which is quenched along a control path subject to noise. The Hamiltonian is
\begin{align}\label{eq:isingHam}
    H(h) = -J \sum_{j=1}^L \left( \sigma_j^x \sigma_{j+1}^x + h \sigma_j^z \right)
\end{align}
 where $J$ is the intrinsic energy scale governing the nearest neighbour interactions of the system. This setup is both analytically tractable and experimentally realisable in platforms such as trapped ions \cite{Jurcevic2014TrappedIons}, Rydberg arrays \cite{Borish2020,Scholl2021}, and superconducting qubits \cite{Guo2019SuperconductingQ}.

Following an approach developed in \cite{quantumWorkStats} (see Appendix~\ref{app:ising} for details), in the limit of large $N$ and $L\rightarrow \infty$, the stochastically averaged dissipated work takes the form
\begin{align}\label{eq:IsingWdiss}
    \myinnerp*{W_\text{diss}} =  L\int_0^1 dt \ \left(\frac{1}{N}\Dot{h}_t^2 + N \Phi_t \right)  \int^\pi_0 \frac{dk}{4\pi}C(k,h_t)
\end{align}
which shows that the dissipated work per site remains finite in the thermodynamic limit. The function $C(k,h)$ is defined as
\begin{align}\label{eq:C(k,h)}
 C(k,h)&=2 \beta \tfrac{J^{4}}{\epsilon_k^2}
\big(h-\cos k\big)^{2}\,\sech^{2}\big(\beta \epsilon_{k}\big) \nonumber \\
& \ \ \ \ \ \ \ \ \ \  \ \ \ \ +2 \tfrac{J^{4}}{\epsilon_k^3}\sin^{2}(k)\tanh\big(\beta \epsilon_{k}\big)
\end{align}
where $\epsilon_k =J\sqrt{h^2 + 1 -2 h \cos(k)}$. 

It is known that a phase transition between ferromagnetic ($\abs{h}<1$) and paramagnetic ($\abs{h}>1$) states occurs sharply at the critical point $h=1$ in the zero-temperature limit; this transition is broadened as temperature increases \cite{Sachdev_2011}. The inset of Fig.~\ref{fig:isingVsPhi} shows the magnetisation along the $z$-component for different values of $h$. At lower temperatures, the spin-$z$ expectation per site $\langle \sigma_z \rangle$ approaches $1$ as $h$ increases beyond $J$, and approaches $0$ in the absence of a transverse field. Since thermal fluctuations counteract ferromagnetic alignments, as $\beta\rightarrow 0$, the magnetisation becomes insensitive to control over $h$. By the same token, strong magnetisations $\langle \sigma_z \rangle \approx \pm 1$ require $\abs{h} \gg J$. Here, we choose to look at the Landauer erasure of the Ising chain, sweeping from $h_0=0.0$ to a defined $h_1$, destroying long-range order upon entering the paramagnetic phase.

As in Fig.~\ref{fig:qubit}, we again found $W_\text{diss}$ to agree with fully numeric calculations and to have a minimum of $W_\text{opt}$ at optimal duration $N_\text{opt}$. We are primarily interested in the effects of changing temperature and noise strength on $N_\text{opt}$ and $W_\text{opt}$. (In Appendix~\ref{app:ising} we find that $N_\text{opt}$ and $W_\text{opt}$ are largely independent of $L$, beyond $L>4$). We keep $J=1$, and define other energy-related quantities, such as $\beta$, in relation to $J$.

In Fig.~\ref{fig:isingVsPhi}, we see that for all noise strengths, $N_\text{opt}$ is reduced as $\beta$ increases,  changing most noticeably around $\beta \approx 1$. Therefore, it is beneficial to drive the protocol faster at lower temperatures. However, when temperatures are lowered, we can see that the average dissipated work monotonically increases, rising to a maximum value at $\beta \approx 1$. Thus, at larger temperatures, where the linear response framework applies most comfortably, $W_\text{opt}$ is minimised and $N_\text{opt}$ has the useful property of being largely insensitive to the exact temperature. 



\begin{figure}[t]
    \centering
    \includegraphics[clip, trim=0cm 0.1cm 0cm 0cm, width=\linewidth]{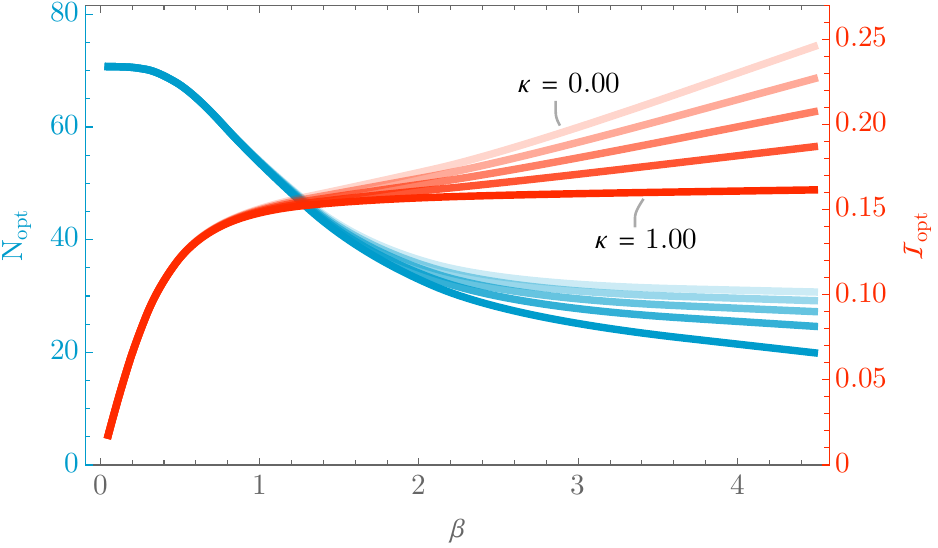}
    \caption{The interpolated cost function $\mathcal{I}_\text{opt}(\kappa)$ (red) increasing with $\beta$ in units of $[\Delta]$ and associated $N_\text{opt}$ (blue) decreasing with $\beta$. Darker colours indicate that $\kappa$ is closer to $1$. Parameter values are identical to those in Fig.~\ref{fig:qubit}.
    \label{fig:NoptIoptQubitPolaron}}
\end{figure}

\textit{Work Fluctuations}---The approach outlined in Eqs.~\eqref{eq:relent}-\eqref{eq:Wopt} can be extended to the variance of dissipated work, $\sigma_W^2=\langle W^2\rangle-\langle W\rangle^2$. Measurements of this quantity are constructed via a series of two-point energy measurement schemes \cite{esposito2009nonequilibrium} across each quench (for discussion on the validity of this implementation, see Appendix~\ref{app:noisyTPMS}). 
After the $N$\textsuperscript{th} quench, under the same linear response assumptions one can obtain a noise-averaged expression for the variance in work (see Appendix~\ref{app:CGF} 
for details):
\begin{align}\label{eq:variance}
    \frac{\beta}{2}\myinnerp*{\sigma_W^2}  \simeq \frac{1}{2}\int^1_0 dt \  \left( \frac{1}{N}  m_{ij}(\vec{v}_t) \Dot{v}_t^i\Dot{v}_t^j + N  m_{ij}(\vec{v}_t)\Phi^{ij}_{t} \right) 
\end{align}
where 
\begin{align}\label{eq:varMetric}
     m_{ij}(\vec{v}_t)&=\beta \Big(\! \tfrac{1}{2}\Tr\left[ \{V_i, V_j\} \pi_t \right]-\Tr\left[V_i\pi_t\right]\Tr\left[V_j\pi_t\right] \! \Big)
\end{align}
is another metric tensor on the space of control variables. It coincides with $g_{ij}(\vec{v}_t)$ in the classical commuting limit, for example when $[V_j,\pi_t]=0$ for all $j$ and $t$. However, in general the metrics are not equivalent if the Hamiltonian does not commute at different points along the protocol, leading to an effect known as quantum friction \cite{PhysRevLett.123.230603}.


The fact that, in general, $m_{ij}(\vec{v}_t) \neq g_{ij}(\vec{v}_t)$ means that the duration $N_\text{opt}$ from Eq.~\eqref{eq:Nopt} will not minimise the work fluctuations. To see this difference in optimal steps one can define a cost function $\mathcal{I}(\kappa) := \kappa \myinnerp*{W_{\text{diss}}} + (1-\kappa ) \frac{\beta}{2} \myinnerp*{\sigma_W^2}$ that interpolates between the average dissipation and its fluctuations with parameter $\kappa\in[0,1]$ \cite{erdman2023pareto}. This minimum interpolated cost, $\mathcal{I}_\text{opt}(\kappa)$, and its optimal number of steps $N_{\text{opt}}(\kappa)$ are presented in Fig.~\ref{fig:NoptIoptQubitPolaron}. We see that at lower temperatures, where quantum friction effects are less suppressed by thermal influences, the $\kappa$ dependence becomes more significant both for $N_\text{opt}(\kappa)$ and $\mathcal{I}_\text{opt}(\kappa)$.


\textit{Conclusion}---For stochastic quantum step-equilibration processes, we predict the existence and magnitude of an optimal number of steps and a corresponding minimum dissipation quantified by thermodynamic length. Our formalism applies to weak noise processes within the linear response regime. This prediction overturns the intuition that 
``the most energy efficient protocols are the slowest'', and so has direct impact on optimising energetic efficiencies of a wide variety of quantum machines. Furthermore, the only additional information needed to estimate $N_\text{opt}$ and $W_\text{opt}$ is the variance of noise increments $\Phi_t^{ij}$ for each control parameter. 

This work can be extended to look at autonomous quantum systems, where entropic biases required to drive control parameters are included in the thermodynamic description of a protocol or complete cycle. Such cases may arise in classical-quantum hybrid systems \cite{layton2025restoringsecondlawclassicalquantum,eglinton2024stochastic} where back-action effects must also be considered. It would also be of interest to investigate any repercussions of our methodology and findings beyond the close-to-equilibrium regime, for example, in deriving more realistic noise-dependent quantum thermodynamic speed limits \cite{van2021geometrical}.  
\\

\acknowledgments

This work was supported by the Engineering and Physical Sciences Research Council [grant number EP/W524347/1]. H.J.D.M. acknowledges funding from a Royal Society Research Fellowship (URF/R1/231394). We thank Katarzyna Macieszczak for helpful discussions.
The data that support the findings of this article are openly available \cite{data}.

\bibliography{apssamp}


\clearpage
\onecolumngrid
\appendix
\setcounter{secnumdepth}{1}


\section{Noise expansions}
\label{app:noise}

In this appendix, we start from the established deterministic expression for $W_\text{diss}$ \cite{quantumWorkStats} (which underpins Eq.~\eqref{eq:length} and is derived by substituting into Eq.~\eqref{eq:relent} the 1\textsuperscript{st} order approximated evolution of $\pi_n$ between subsequent steps \cite{scandi2018quantifying}),
\begin{align}
    W_\text{diss} = \frac{\beta }{2} \sum_{n=0}^{N-1} \Tr [dH_n \mathds{J}_{\pi_n} \left[ \Delta_{\pi_n}(dH_n) \right] ] \label{eq:Wdiss},
\end{align}
and seek to find its closed form modification in the case of stochastic control amplitudes. This is achieved via a 2\textsuperscript{nd} order expansion in noisy parameters, $\Tilde{H}_n$ and $d\Tilde{H}_n$. We are first interested in deriving an expansion for the system Gibbs state. This result serves as the starting point for further expansions—most notably the expansion of the Kubo covariance $\mathds{J}_{\pi_n}[\Delta_{\pi_n} (A)]$
where
\begin{align}\label{eq:Jpi}
    \mathds{J}_{\rho} \left[ A \right] := \int^1_0 \rho^x A \rho^{1-x} dx
\end{align}
and $\Delta_\rho(A) := A - \Tr (A \rho)$.
This, in turn, is used to calculate potential corrections to the metric $g^{(n)}_{ij} = \beta\Tr[ V^i \mathds{J}_{{\pi}_n}[ \Delta_{{\pi}_n}(V^j)]]$ (c.f. Eq.~\eqref{eq:KuboMetric}) and, after the appropriate noise-averaging, culminates in a modified formula for the dissipated work \eqref{eq:WdissGeo}. The same treatment can be applied to the full cumulant generating function (as given by Eq.~\eqref{eq:Kdiss} in Appendix~\ref{app:CGF}), but we shall here remain focussed on $W_\text{diss}$.

Using the Dyson series for exponential operators \cite{hiai2014introduction}, 
we can expand a matrix exponential about $A$ in a small perturbation $tB$ where $t$ is a small parameter and $A$ and $B$ are, in general, non-commuting matrices:
\begin{align}\label{eq:Exp(A+tB)}
    e^{A+tB}=\sum_{k=0}^\infty t^k A_k
\end{align}
where for $s \in \mathds{R}$
\begin{align}\label{eq:Exp(A+tB)2}
    A_0 & = e^{A}, \nonumber \\
    A_k & = \int^1_0 dt_1 \int^{t_1}_0 dt_2 ...\int^{t_{k-1}}_0 dt_k ~e^{(1-t_1)A} B e^{(t_1-t_2)A} B~ ...~ B e^{t_k A}. 
\end{align}

To expand Eq.~\eqref{eq:Wdiss} to 2\textsuperscript{nd} order in noise, we require an adapted expression for the Gibbs state, $\pi_n \mapsto \pi_n + \delta^{(1)}_{\pi_n} + \delta^{(2)}_{{\pi}_n} + \mathcal{O}(\Tilde{H}_n^3)$ where $\pi_n = e^{-\beta H_n}/Z_n$ is the deterministic state, $\delta^{(1)}_{{\pi}_n} $ ($\delta^{(2)}_{{\pi}_n} $) is a traceless 1\textsuperscript{st} (2\textsuperscript{nd}) order superoperator in $\Tilde{H}_n$, and higher order terms $\mathcal{O}( \Tilde{H}_n^3 )$ are neglected.
In relation to the thermal states of the quench process, we have to 2\textsuperscript{nd} order,
\begin{align}\label{eq:Exp(H)_expansion}
    e^{-\beta( {H}_n+\Tilde{H}_n)} =& ~ e^{-\beta  {H}_n} \nonumber \\ 
    & -\beta \int^1_0 dx ~e^{-\beta {H}_n(1-x)} \Tilde{H}_n e^{-\beta {H}_n x} \nonumber \\
    & + \beta^2 \int^1_0 dx \int^x_0 dy ~e^{-\beta {H}_n(1-x)} \Tilde{H}_n e^{-\beta {H}_n (x-y)} \Tilde{H}_n e^{-\beta {H}_n y}.
\end{align}
Plugging this expansion into $\pi_n = e^{-\beta( {H}_n+\Tilde{H}_n)}/\Tr[e^{-\beta( {H}_n+\Tilde{H}_n)}]$, and performing a binomial expansion of the denominator (i.e., the partition function) of the form $(1+c)^{-y} \approx 1-yc+\frac{y(y+1)}{2}c^2$, we get \cite{scandi2018quantifying}
\begin{align}\label{eq:pi_expansion}
    \pi_n \mapsto & ~ {\pi}_n  -\beta \mathds{J}_{{\pi}_n}[ \Delta_{ {\pi}_n}(\Tilde{H}_n)] + \beta^2 \mathds{K}_{ {\pi}_n}[\Tilde{H}_n]
\end{align} where $\mathds{J}_{ {\pi}_n}[ \Delta_{ {\pi}_n}(\Tilde{H}_n)]$ and $\mathds{K}_{ {\pi}_n}[\Tilde{H}_n]$ are indeed traceless and we can identify the correction terms as $\delta^{(1)}_{\pi_n} (\Tilde{H}_n)  = -\beta \mathds{J}_{ {\pi}_n}[ \Delta_{ {\pi}_n}(\Tilde{H}_n)]$ and $ \delta^{(2)}_{{\pi}_n} (\Tilde{H}_n)  = \beta^2 \mathds{K}_{ {\pi}_n}[\Tilde{H}_n]$. The 2\textsuperscript{nd} order superoperator takes the form
\begin{align}\label{eq:K_expansion}
    \mathds{K}_{\rho}[A]= - \Tr\left[ \rho A \right] \mathds{J}_{\rho}\left[ \Delta_{\rho}(A) \right] + \rho \Delta_{\rho}\left( \int^1_0 du \int^u_0 dy~ \rho^{-u} A \rho^{(u-y)} A \rho^{y}\right).
\end{align}

With this noise expansion of the Gibbs state, we can calculate the remaining ingredients needed to update Eq.~\eqref{eq:Wdiss}: that is, expansions of $\Delta_{\pi_n} (A)$, $\pi_n^x$ and then finally of $\mathds{J}_{\pi_n}[\Delta_{\pi_n} (A)]$ in powers of $\Tilde{H}_n$
\begin{align} \label{eq:JpiExpansion}
    \mathds{J}[\Delta(A)] \mapsto\mathds{J}[\Delta (A)] + \delta^{(1)}_{\mathds{J}[\Delta]}(A) + \delta^{(2)}_{\mathds{J}[\Delta]}(A) + \mathcal{O}(\Tilde{H}_n^3)
\end{align}
where subscripts of $\pi_n$ have been temporarily suppressed.


For the calculation of $W_\text{diss}$ we also require an expansion of powers of Gibbs states, which turns out to be 
\begin{align}\label{eq:pi^x_expansion}
    \pi_n^x \mapsto & ~  {\pi}_n^x -\beta \mathds{J}_{ {\pi}_n}^{(x)} [ \Delta_{ {\pi}_n}(\Tilde{H}_n) ] + \beta^2 \mathds{K}_{ {\pi}_n}^{(x)} [\Tilde{H}_n]
\end{align}
where $\mathds{J}_{\rho}^{(x)} \left[ A \right] = \int^x_0 \rho^u A \rho^{x-u}du$ and $\mathds{K}_{\rho}^{(x)} \left[ A \right]$ has the non-trivial form:
\begin{align}\label{eq:K^x_expansion}
    \mathds{K}_{\rho}^{(x)}[A] =& x^2 \int^1_0 du \int^u_0 dy~ \rho^{x(1-u)} A \rho^{x(u-y)} A \rho^{xy}  \nonumber \\
    & - x^2 \mathds{J}_{\rho}^{(x)} \left[ A \right] \Tr \left[ \mathds{J}_{\rho} \left[ A \right] \right]\nonumber \\
    & - x \rho^x \Tr \left[ \int_0^1 du \int_0^u dy~ \rho^{1-u}A\rho^{u-y}A \rho^y \right] \nonumber \\
    & + \frac{x(x+1)}{2} \rho^x \Tr\left[ \mathds{J}_{\rho} \left[ A \right] \mathds{J}_{\rho} \left[ A \right] \right].
\end{align}
The last expansion in our required toolkit is 
\begin{align}\label{eq:Deltapi_expansion}
    \Delta_{\pi_n}(A) \mapsto \Delta_{ {\pi}_n}(A) + \Tr \left[ A \beta \mathds{J}_{ {\pi}_n} [ \Delta_{ {\pi}_n} (\Tilde{H}_n) ]
    -A \beta^2 \mathds{K}_{ {\pi}_n} [ \Tilde{H}_n]\right] +\mathcal{O}(\Tilde{H}_n^3).
\end{align}


We now have all the necessary definitions to expand Eq.~\eqref{eq:Jpi} from which we will calculate the average dissipated work. Substituting in Eqs.~\eqref{eq:pi_expansion}, \eqref{eq:pi^x_expansion}, and \eqref{eq:Deltapi_expansion}, we find the full form of the terms in Eq.~\eqref{eq:JpiExpansion},
$\mathds{J}[\Delta(A)] \mapsto\mathds{J}[\Delta (A)] + \delta^{(1)}_{\mathds{J}[\Delta]}(A) + \delta^{(2)}_{\mathds{J}[\Delta]}(A) + \mathcal{O}(\Tilde{H}_n^3)$:
\begin{align} \label{eq:deltaJ1}
    \delta^{(1)}_{\mathds{J}[\Delta]}(A) = & \beta \pi_n \Tr \left\{ A \mathds{J}_{\pi_n} [ \Delta_{\pi_n} (\Tilde{H}_n) ] \right\}  \nonumber \\ 
    &- \beta \int_0^1 \mathds{J}_{\pi_n}^{(1-x)} [\Delta_{\pi_n}(\Tilde{H}_n)] \Delta_{\pi_n}(A) \pi_n^x 
    + \pi_n^{1-x} \Delta_{\pi_n}(A) \mathds{J}_{\pi_n}^{(x)}[\Delta_{\pi_n}(\Tilde{H}_n)] ~dx
\end{align}
and
\begin{align} \label{eq:deltaJ2}
    \delta^{(2)}_{\mathds{J}[\Delta]}(A) = & - \beta^2 \pi_n \Tr \left\{ A \mathds{K}_{\pi_n} [\Tilde{H}_n ] \right\} \nonumber \\ 
    &+ \beta^2 \int_0^1 \mathds{K}_{\pi_n}^{(1-x)}[\Tilde{H}_n] \Delta_{\pi_n}(A) \pi_n^x 
    + \pi_n^{1-x} \Delta_{\pi_n}(A) \mathds{K}_{\pi_n}^{(x)}[\Tilde{H}_n] ~dx\nonumber \\
    &+ \beta^2 \int_0^1 \mathds{J}_{\pi_n}^{(1-x)}[\Delta_{\pi_n}(\Tilde{H}_n)]\Delta_{\pi_n}(A) 
    \mathds{J}_{\pi_n}^{(x)}[\Delta_{\pi_n}(\Tilde{H}_n)] ~dx \\
    & - \beta^2 \Tr \left\{ A \mathds{J}_{\pi_n} [ \Delta_{\pi_n} (\Tilde{H}_n) ] \right\} \int_0^1\pi_n^{1-x}\mathds{J}_{\pi_n}^{(x)}[\Delta_{\pi_n}(\Tilde{H}_n)]
    + \mathds{J}_{\pi_n}^{(1-x)}[\Delta_{\pi_n}(\Tilde{H}_n)]\pi_n^{x}~dx \nonumber
\end{align}
where $\delta^{(1)}_{\mathds{J}[\Delta]}(A) \in \mathcal{O}(\Tilde{H}_n)$ and $\delta^{(2)}_{\mathds{J}[\Delta]}(A) \in \mathcal{O}(\Tilde{H}_n^2)$.

With this information, we expand Eq.~\eqref{eq:Wdiss}
as
\begin{align}\label{eq:WdissExpand}
    \myinnerp*{W_\text{diss}} \mapsto \frac{\beta}{2} \sum_{n=0}^{N-1} \myinnerp*{ \Tr \left[ (dH_n + d\Tilde{H}_n ) \left\{ \mathds{J}_{\pi_n}[\Delta_{\pi_n}(dH_n + d\Tilde{H}_n)] 
    + \delta^{(1)}_{\mathds{J}[\Delta]}(dH_n + d\Tilde{H}_n)
    + \delta^{(2)}_{\mathds{J}[\Delta ]}(dH_n + d\Tilde{H}_n)
    \right\} \right] }.
\end{align}
Using the property that both corrections are linear in their argument such that $\delta^{(i)}_{\mathds{J}[\Delta]}(A+B) = \delta^{(i)}_{\mathds{J}[\Delta]}(A)+\delta^{(i)}_{\mathds{J}[\Delta]}(B)$, we expand in orders of noise
\begin{align}\label{eq:WdissExpand2}
    \myinnerp*{W_\text{diss}} = \frac{\beta}{2} \sum_{n=0}^{N-1} & \Tr \left[ dH_n \mathds{J}_{\pi_n}[\Delta_{\pi_n}(dH_n)] \right] \nonumber \\
    + & \myinnerp*{ \Tr \left[ dH_n \mathds{J}_{\pi_n}[\Delta_{\pi_n}(d\Tilde{H}_n)] \right] + \Tr \left[ d\Tilde{H}_n \mathds{J}_{\pi_n}[\Delta_{\pi_n}(dH_n)] \right] + \Tr \left[ dH_n \delta^{(1)}_{\mathds{J}[\Delta]} (dH_n) \right]  } \nonumber \\
    + & \myinnerp*{\Tr \left[ d\Tilde{H}_n \mathds{J}_{\pi_n}[\Delta_{\pi_n}(d\Tilde{H}_n)] \right] +\Tr \left[ d\Tilde{H}_n \delta^{(1)}_{\mathds{J}[\Delta]} (dH_n) \right] + \Tr \left[ dH_n \delta^{(1)}_{\mathds{J}[\Delta]} (d\Tilde{H}_n) \right] + \Tr \left[ dH_n \delta^{(2)}_{\mathds{J}[\Delta]} (dH_n) \right]  
    } \nonumber \\
    + & \mathcal{O}\big((\Tilde{H}_n,d\Tilde{H}_n)^3\big).
\end{align}
This equation provides complete information about the dissipated work for a driven, near-equilibrium quantum system, with control noise incorporated up to second order. Any further simplification requires additional approximations or assumptions about the specific noise processes involved, which we will employ.

In Eq.~\eqref{eq:WdissExpand2}, the term in the first line reproduces the noiseless result \eqref{eq:Wdiss}. Terms in the second line vanish for zero-mean Gaussian noise processes, as they are $1$\textsuperscript{st} order in noise. Out of the $2$\textsuperscript{nd} order noise terms in the third line, by an order of magnitude the largest contribution comes from $d\Tilde{H}_n \mathds{J}_{\pi_n}[\Delta_{\pi_n}(d\Tilde{H}_n)]$. As briefly stated in the main text—and now made explicit by the above derivation—all remaining contributions appear only at order $\mathcal{O}(\epsilon^3)$ and higher, where $\epsilon \in \{ dH_n, d\Tilde{H}_n, \Tilde{H}_n \}$. We will continue to ignore the higher order noise terms in the last line.

Disregarding the $1$\textsuperscript{st} order noise terms, we then decompose the Hamiltonian contributions according to Eq.~\eqref{eq:Ham} which defers noise-averaging into the relevant correlation functions,
\begin{align}
    \myinnerp*{W_\text{diss}}  & = \frac{\beta}{2}\sum_{n=0}^{N-1} \Big\{ dv^\alpha_n \Tr[ V^\alpha \mathds{J}_{ {\pi}_n}[ \Delta_{ {\pi}_n}(V^\beta) ]] dv^\beta_n 
    + \Phi_{nn}^{\alpha\beta} \Tr[ V^\alpha \mathds{J}_{ {\pi}_n}[ \Delta_{ {\pi}_n}(V^\beta) ]] \Big\} + \beta \eta^\Psi_{W} + \beta^2 \eta^\chi_{W} . \label{eq:W_dissDecomp} 
\end{align}
where the first summation translates to the final discrete expression \eqref{eq:WdissGeo} from which the other main results of this work are derived.

In Eq.~\eqref{eq:W_dissDecomp}, the small corrections ignored in the main text are
\begin{align}\label{eq:WdissCorrections}
    \eta^\Psi_{W} & =\frac{\beta}{2}\sum_{n=0}^{N-1}
     \myinnerp*{\Tr \left[ d\Tilde{H}_n \delta^{(1)}_{\mathds{J}[\Delta]} (dH_n) \right] + \Tr \left[ dH_n \delta^{(1)}_{\mathds{J}[\Delta]} (d\Tilde{H}_n) \right] } \\
    \eta^\chi_{W} & =\frac{\beta}{2}\sum_{n=0}^{N-1} \myinnerp*{\Tr \left[ dH_n \delta^{(2)}_{\mathds{J}[\Delta]} (dH_n) \right]  
    }.\label{eq:WdissCorrections2}
\end{align}
These 
corrections follow from the remaining $2$\textsuperscript{nd} order noise terms in Eq.~\eqref{eq:WdissExpand2}.

The superscripted labels $\Psi$ and $\chi$ refer to the correlation functions $\Psi^{\alpha \beta}_{n m}$ and $\chi^{\alpha \beta}_{n m}$ that feature in their definitions. These correlation functions are worth defining:  $\chi^{\alpha \beta}_{n m} = \myinnerp*{ \xi_n^\alpha \xi_m^\beta}$ accounts for the variance of, and instantaneous covariance between, each total noise process; and correlations between the entire history of each process and its next increment are described by $\Psi^{\alpha \beta}_{n m} = \myinnerp*{\Delta\xi_n^\alpha \xi_m^\beta } = \sum_{i=1}^{m-1} \myinnerp*{ \Delta\xi_n^\alpha \Delta\xi_i^\beta } = \sum_{i=1}^{m-1} \Phi^{\alpha \beta}_{n i}$ using the fact that $\xi_m^\beta = \sum_{i=1}^{m-1} \Delta\xi_i^\beta$. 
For non-interacting noise sources, $\chi^{\alpha \beta}_{n m} = \delta^{\alpha\beta} \chi^\alpha _{nm}$ where $\chi^\alpha _{nm}$ is the noise \textit{autocorrelation} function and $\delta$'s represent Kronecker deltas – the same applies for $\Psi^{\alpha \beta}_{n m}$ and $\Phi^{\alpha \beta}_{n m}$. Since only single-time correlation functions feature in Eq.~\eqref{eq:W_dissDecomp}, we can drop the double subscript, such that $\chi^{\alpha \beta}_n = \myinnerp*{ \xi_n^\alpha  \xi_n^\beta}$ and similarly for $\Psi^{\alpha \beta}_n$ and $\Phi^{\alpha \beta}_n$.

\section{Alternative Markovian Gaussian noise processes}
\label{app:markov}

Here, we return our attention to Markovian Gaussian noise sources. There exist (at least) three established models: Gaussian white noise (GWN), Wiener, and Ornstein-Uhlenbeck (O-U) processes \cite{Levy2020}. We aim to investigate how each of these processes contribute to thermodynamic performance following the approach laid out in the main body. Numeric simulations are then given for the example of a driven qubit. 

In our analysis, noise is implemented discretely. So, we must define the discrete analogue to the above processes--- for the O-U process it is the first-order autoregressive model, AR(1) \cite{shumstof2025}. All these processes can be constructed using independent GWN increments, $\Delta\eta$:
\begin{align}
    \text{GWN:}~~~& \xi_{n+1} = \Delta\eta_n \label{eq:processGWN} \\
    \text{Wiener:}~~~& \xi_{n+1} =\xi_{n}+\Delta\eta_n \label{eq:processWiener}\\
    \text{AR(1):}~~~& \xi_{n+1} = \phi \xi_n + \Delta\eta_n \label{eq:processAR1}
\end{align}
where $0 \leq \phi \leq 1 $ is the response function interpolating between GWN and Wiener processes. Simply by observing that $\myinnerp*{\Delta\eta_n} = 0$, we can see that all these processes are indeed zero-mean. From \eqref{eq:processGWN}-\eqref{eq:processAR1}, we can calculate the variance of noise increments for each process, $\Phi_n = \myinnerp*{ \Delta\xi_n \Delta\xi_n }$:
\begin{align}
    \Phi_n^\text{GWN }&  = 2\sigma_\eta^2 \label{eq:varGWN} \\
    \Phi_n^\text{Wiener} & =\sigma_\eta^2\label{eq:varWiener}\\
    \Phi_n^\text{AR(1)} & = \frac{(\phi-1)^2}{1-\phi^2} \sigma_\eta^2 + \sigma_\eta^2 .\label{eq:varAR1}
\end{align}
where the variance of GWN increments is $\sigma_\eta^2 = \myinnerp*{ \Delta\eta_n \Delta\eta_n }$.

\begin{figure}
    \centering
    \includegraphics[width=0.5\linewidth]{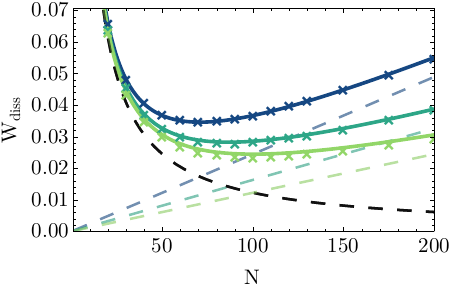}

    \caption{$W_\text{diss} ~[\Delta]$ against number of protocol steps $N$ for different zero-mean Gaussian noise sources. Dark blue lines (and darker grey dotted lines) represent results for GWN, turquoise (and medium-grey dotted lines) for AR(1) process with $\phi=0.5$, and lighter green (and lighter grey dotted lines) for Wiener process. The noiseless contribution exhibits the same $1/N$ behaviour in all cases, while the noise-induced $N$-scaling contribution differs by a factor of two between GWN and Wiener processes. Parameters: $\beta =0.1$, $\alpha=0.0005$, $\sigma_\eta = 0.05$.}
    \label{fig:AltNoise}
\end{figure}

Given that to $2$\textsuperscript{nd} order all significant thermodynamic influence from noise comes via $\Phi_n$, we can make a trivial prediction about the effects of changing noise processes: the second term in \eqref{eq:WdissCont} is twice the magnitude for GWN processes than for Wiener processes. This entails that $W_\text{opt}$ is a factor of $\sqrt{2}$ larger for GWN processes than for Wiener processes and the reciprocal is true for $N_\text{opt}$ as per \eqref{eq:Nopt}. Meanwhile, equivalent effects of AR(1) processes exist between these two extremes depending on $\phi$. Of course, to make such comparisons, we assume the $\{\Delta\eta_n\}$ used to construct each process obey the same statistics. These predictions are displayed in Fig.~\ref{fig:AltNoise}.

Next, we briefly comment on the constraints on the domain of valid noise parameters for GWN and Wiener processes in our approach. For Wiener noise, the 2\textsuperscript{nd} order weak noise approximation $\pi_n \mapsto \pi_n + \delta^{(1)}_{\pi_n} + \delta^{(2)}_{{\pi}_n} + \mathcal{O}(\Tilde{H}_n^3)$, valid under the condition $\norm*{\Tilde{H}_n}  \ll \norm*{ H_n }$, begins to break down sooner than its GWN counterpart. This is because the variance of $\Tilde{H}_n$ grows linearly with $N$ for Wiener processes, meaning some trajectories may start to breach this condition. 
In contrast, from \eqref{eq:varGWN} and \eqref{eq:varWiener}, for a GWN noise process, the variance of $dH_n$ is twice that of its equivalent Wiener process. This means that, as parameter magnitudes are pushed to their limits, the linear response condition, $\beta \norm*{d\Tilde{H}_n + dH_n}  \ll 1$, may be breached sooner by some GWN trajectories.

Returning to Eqs.~\eqref{eq:WdissCorrections} and \eqref{eq:WdissCorrections2}, looking at the dependence of $\eta^\Psi_{W}$ and $\eta^\chi_{W}$ on $dH_n$, $d\Tilde{H}_n$ and $\Tilde{H_n}$, one can deduce their $N$-scaling behaviour. For GWN, where covariances are $n$-independent, $\eta^\Psi_{W} \propto N^0 (const.)$ and $\eta^\chi_{W} \propto 1/N$. Since neither term grows with $N$, we are safe to neglect their contributions even in the large $N$ limit. For Wiener noise, the autocorrelation function can be calculated under the present indexing convention as $\chi_n^\text{Wiener} = n \Phi_n^\text{Wiener}$ and $\Psi_n^\text{Wiener}$ vanishes. This means that the correction $\eta^\Psi_{W}$ also vanishes. Since $\chi_n^\text{Wiener}$ is no longer constant, using $n=Nt$, the $N$-scaling of the $\chi$-related correction becomes $\eta^\chi_{W} \propto N^0 (const.)$. Therefore, corrections \eqref{eq:WdissCorrections} and \eqref{eq:WdissCorrections2} do not grow with $N$, and so remain subleading in the large-$N$ analysis.\\

\begin{figure}[ht]
    \centering
    \subfloat[]
    {\includegraphics[width=0.48\columnwidth]{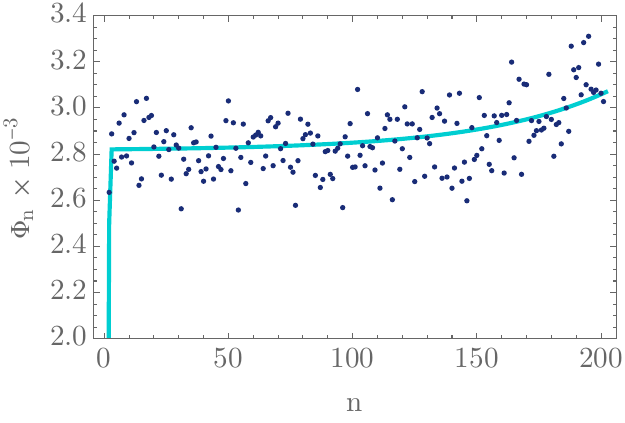}
    \label{fig:AR(n)_Phi}
    }
    \hfill
    \subfloat[]{\includegraphics[width=0.49\columnwidth]{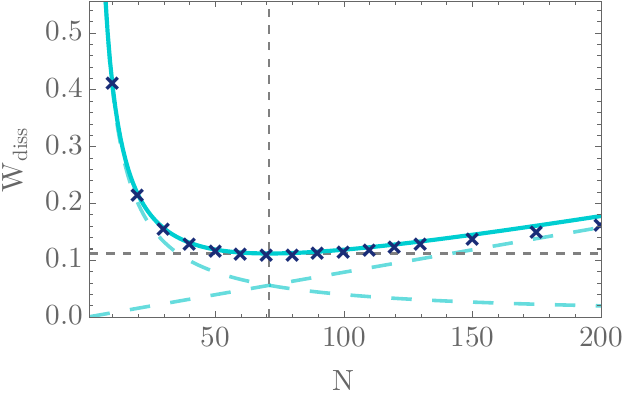}
    \label{fig:AR(n)_Wdiss}
    }
  \caption{(a) The predicted incremental AR($n$) process variance, $\Phi_n^{\text{AR(}n{)}}$, as a discrete function of $n$, plateauing quickly after  $n > 2$ before slowly diverging. (b) The average dissipated work for the same AR($n$) process with crosses, dashed and dotted lines indicating the same quantities as in Fig.~\ref{fig:qubit}. Though not obvious, the increasing dotted line is in fact non-linear, according to the behaviour of $\Phi_n^{\text{AR(}n{)}}$ with $n$. Parameters: $r=500$, $\beta=1$, $\sigma_\eta = 0.05$, $\lambda_\eta = 1$, 
  $A=0.642$, $\alpha =0.5$, $\Delta =1$, $\omega_0=-5$, $\omega_1=5$.} 
\end{figure}

The formulation of the AR(1) process in Eq.~\eqref{eq:processAR1} can be extended to non-Markovian sources; namely, the $p$\textsuperscript{th}-order AR($p$) noise processes defined by a set of response functions $\phi_i$, 
\begin{align}
    \text{AR(p):}~~~& \xi_{n+1} = 
    \sum_{i=1}^p \phi_i \xi_{n+1-i} + \Delta\eta_n .
    \label{eq:processAR(n)}
\end{align}
Here, we will consider the case where memory of the entire preceding $n$ steps is retained ($p = n \geq 1$). It is useful to define influence-response coefficients,  
\begin{align}
    h_{n} = \sum_{i=1}^n \phi_i h_{n-i} .\label{eq:h(n)}
\end{align}
We are then required to set initial values, which we choose to be $h_{-1} = 0$ and $h_0 = 1$.

With these quantities established, we derive the variance in AR($n$) increments to be
\begin{align}
    \Phi_n^{\text{AR(}n{)}} & = \sigma_\eta^2 \sum_{m=0}^n (h_m - h_{m-1})^2.
    \label{eq:varAR(n)}
\end{align}
We are free to chose values for the $\phi_i$ that, together with $\sigma_\eta^2$, define the statistics. As an example, we chose to look at systems with memories that exponentially decay, such that $\phi_m = A e^{-\lambda_\eta m}$. The $\Phi_n^{\text{AR(}n{)}}$ behaviour that arises is shown in Fig.~\ref{fig:AR(n)_Phi}, where we have purposely chosen parameters that give rise to an unstable process, diverging at larger $n$. Since, for many pairings of $A$ and $\lambda_\eta$, $\Phi_n^{\text{AR(}n{)}}$ converges and quickly becomes effectively independent of $n$, we chose a diverging setup to emphasize the effects of $n$-dependence in $\Phi_n^{\text{AR(}n{)}}$ and the ability of our approach in capturing such features. When working in the continuous-time limit as in Eq.~\eqref{eq:WdissCont}, we define $\Phi^{\mathrm{AR}(n)}(t)$ as an interpolation of the discrete sequence $\{\Phi_n^{\mathrm{AR}(n)}\}_{n=1}^N$ at the grid points $t_n = n\Delta t$, thereby applying the variance function to $t \in [0,1]$.

In Fig.~\ref{fig:AR(n)_Wdiss}, we see that our analytic calculations from Eqs.~\eqref{eq:WdissCont}, \eqref{eq:Nopt} and \eqref{eq:Wopt} are successful in describing dissipative behaviour for a AR($n$) non-Markov noise model featuring in the controls of a qubit avoided crossing. In general, our predictions and simulations may differ for $\Phi_n^{\text{AR(}n{)}}$ with very strong $n$-dependence. Firstly, for an unstable noise process, more simulations are required to retain its zero-mean property. This means, for a finite number of sample paths, the linear noise terms appearing in the second line of Eq.~\eqref{eq:WdissExpand2} may need to be calculated explicitly. In Fig.~\ref{fig:AR(n)_Wdiss}, the slight difference between the predictions and simulated data points at larger $n$ could be an artefact of this limitation.
Second, $N_\text{opt}$ and $W_\text{opt}$ predictions may also differ from simulated results. This is because, as stated in the main text, predictions \eqref{eq:Nopt} and \eqref{eq:Wopt} rely on the assumption that $\Phi^{\text{AR(}n{)}} \neq \Phi(n)^{\text{AR(}n{)}}$, which is not generally the case. In Fig.~\ref{fig:AR(n)_Wdiss}, these predictions are still accurate because the most significant $n$-dependence is only experienced by protocols that are longer than the optimum duration.




\section{Further Ising study}
\label{app:ising}

\begin{figure}[ht]
    \centering
    \subfloat[]{
    \begin{overpic}[width=0.5\columnwidth]{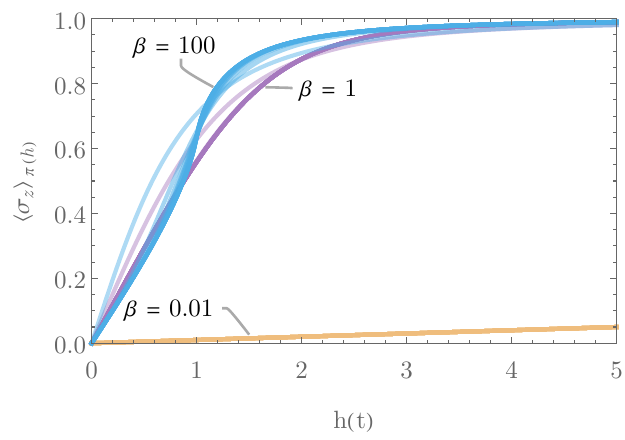}
        \put(50,19){\includegraphics[width=0.225\columnwidth]{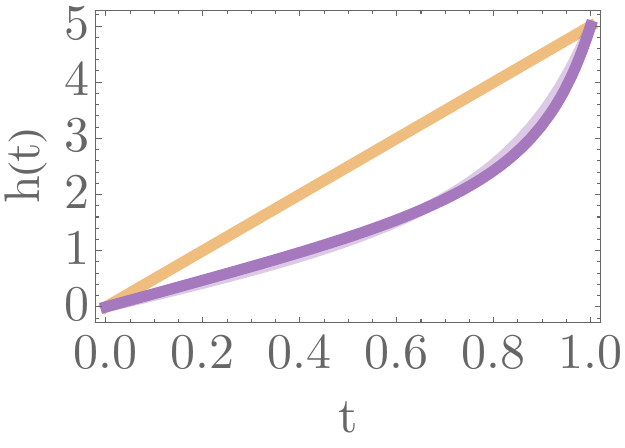}}
    \end{overpic}
    \label{fig:hgeodesic}
    }
    \hfill
    \subfloat[]{\includegraphics[width=0.4\columnwidth]{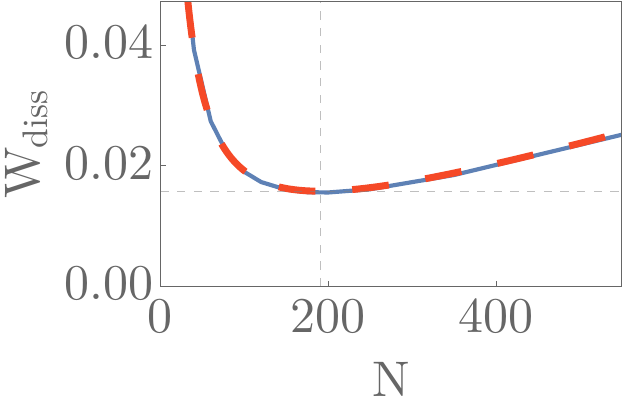}
    \label{fig:isingWdissVsN}
    }
  \caption{(a) The magnetisation per spin across $h$ ($h_1=5.0$) for different temperatures $\beta=0.01$ (orange), $\beta=1.0$ (purple) and $\beta=100$ (blue). Darker lines indicate longer chain lengths and the most transparent lines represent $L=2$. Inset: The optimal path $h(t)$ over time, showing primary dependence is on $\beta$ ($\beta=100$ breaches linear response conditions so its geodesic is not solved) and marginally on $L$. (b) $W_\text{diss}$ in units of $[J]$ for different durations, illustrating the legitimacy of the provided analytic treatment of noise effects --- the blue line representing fully numeric calculations --- and the existence of $W_\text{opt}$ and $N_\text{opt}$ for $L=2$, $\beta = 5$ and $\sigma_n=0.01$.
  }
\end{figure}

In this appendix, we provide more details of our treatment of the TFIM, and further numeric results surrounding the dependence of $N_\text{opt}$ and $W_\text{opt}$ on tunable parameters. First, we expatiate upon the derivation of Eqs.~\eqref{eq:IsingWdiss} and \eqref{eq:C(k,h)} which together provide an equation for $W_\text{diss}$. To do this, and to gather higher order statistics for the dissipated work, we adapt the expression for the cumulant generating function (CGF) (see Eq.~\eqref{eq:KdissDecomp} in Appendix~\ref{app:CGF}) to the TFIM, following closely Appendix G of \cite{quantumWorkStats}.

We begin by performing a Jordan-Wigner transformation \cite{JWtransform,Sachdev_2011} on operators composing the system Hamiltonian:
\begin{align}\label{eq:jordanWigner}
    \sigma_j^z & = 1- 2 c_j^\dagger c_j ,\\
    \sigma_j^+ & = \prod_{i<j} (1-2c_i^\dagger c_i) c_j ,\\
    \sigma_j^- & = \prod_{i<j} (1-2c_i^\dagger c_i) c_j^\dagger .
\end{align}
This transformation maps systems of interacting spin-$1/2$ particles to collections of free fermions with creation and annihilation operators $\{ c_j\}$ that obey canonical anti-commutation rules. We can also choose to work in the momentum basis by replacing the site operators $\{ c_j\}$ with the Fourier decompositions
\begin{align}\label{eq:JWfourier}
    c_j = \frac{1}{\sqrt{L}}\sum_k c_k e^{ijk}.
\end{align}

The transformed Hamiltonian can be expressed as a direct sum over positive $k$-modes, \mbox{$\{k ~|~k= \frac{2\pi (m+1/2)}{L}, ~m=0,1,2,...,\frac{L}{2}-1 \}$},
\begin{align}\label{eq:Hk}
    H(h) & =  \bigoplus_{k>0} H_k(h) \nonumber\\
    & = \bigoplus_{k>0} E_k(h) (c_k^\dagger c_k + c_{-k}^\dagger c_{-k} -1) 
    -i \Omega_k (c_k^\dagger c_{-k}^\dagger + c_k c_{-k})
\end{align}
where $E_k(h) = 2J(h-\cos{k})$ and $\Omega_k=2J\sin{k}$. To be clear, $H_k(h_n)$ denotes the Hamiltonian of the $k$\textsuperscript{th} system mode at the $n$\textsuperscript{th} step.

We have chosen to look at even $L$ and implement anti-periodic boundary conditions at the ends of the chain, such that $c_{j+L} = -c_j$, so there is a clean pairing of $k$ and $-k$ and edge effects do not need special treatment \cite{He_2017}. Note edge effects do not scale with $L$ so, when $L \gg 1$, the bulk dynamics dominate and average properties become equal regardless of chosen boundary conditions. Therefore, at smaller $L$, our results (for average magnetisation, dissipated work etc.) pertain specifically to anti-periodic Ising rings and, as $L$ increases, these results hold for all standard choices of boundary conditions (open or periodic).

\begin{figure}[ht]
    \centering
    \subfloat[]{
        \includegraphics[width=0.47\linewidth]{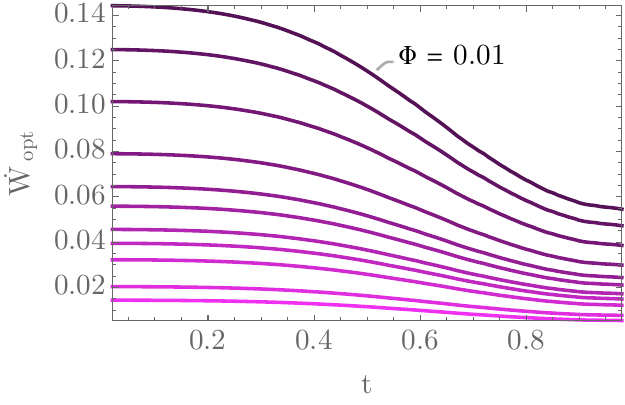}
        \label{subfig:WoptA}
        }
    \hfill
    \subfloat[]{
        \begin{overpic}[width=0.49\linewidth]{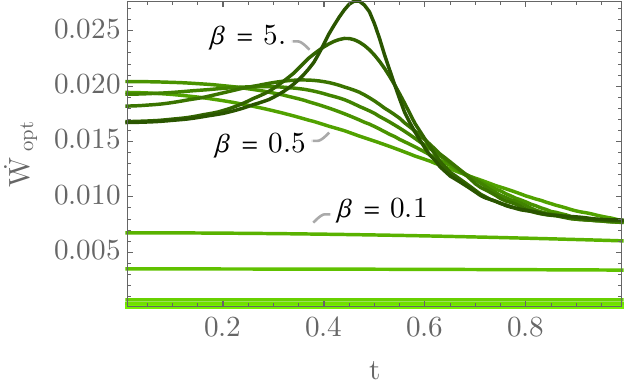}
        \put(66,42){\includegraphics[width=0.16\columnwidth]{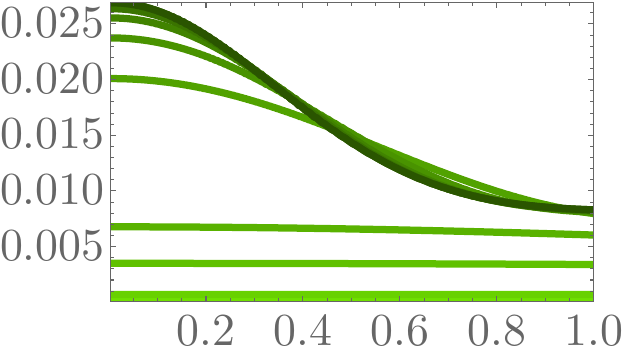}}
        \end{overpic}
        \label{subfig:WoptB}
        }
    \caption{Rate of dissipation against time following the relevant geodesic protocol, in units of $[J^2]$. (a) Noise causes greater dissipation at all times during the stroke, where darker magenta lines denote larger $\Phi \in [ 0.0001,0.01 ]$ and $\beta = 1$. (b) The geodesic path is shown to smooth out the expected dissipative peak around $h_t = 1$ (occurring at $t\approx 0.5$), where darker green lines denote larger $\beta \in [0.001, 10]$ and $\Phi = 0.0002$. Results are for $L = 180$ other that the inset of (b) which is for $L=2$ showing that, since smaller spin chains lack a well-defined phase transition, dissipation is more flat across the protocol.}
    \label{fig:Wopt}
\end{figure}


Next, we decompose the Hamiltonian into independent $(k,-k)$ sectors and work in the fermionic occupation basis $\{\ket{1_k,1_{-k}},\ket{0_k,0_{-k}},\ket{1_k,0_{-k}},\ket{0_k,1_{-k}}\}$. Due to fermion-parity conservation, the Hamiltonian is block diagonal, with a non-trivial $2 \times 2$ block acting on the even-parity subspace. Introducing the Nambu spinor $\Psi_k = (c_k,c_{-k}^\dagger)^T$ \cite{NambuSpinor}, the Hamiltonian for each momentum mode can be written as $H_k(h)= \Psi_k^\dagger \mathtt{H}_k (h) \Psi_k$ where $\mathtt{H}_k(h)$ is the effective $2\times 2$ Bogoliubov–de Gennes Hamiltonian. Following from Eq.~\eqref{eq:Hk}, $H(h)$ has a block-diagonal structure; therefore, $\pi(h)$ is a product state \cite{Sachdev_2011}:
\begin{align}\label{eq:gibbsIsingSupp}
    \pi(h) =\bigotimes\limits_{k>0} \pi_k(h) = \frac{\bigotimes\limits_{k>0} e^{-\beta H_k(h)}}{\prod\limits_{k>0} \Tr[e^{-\beta H_k(h)}]}.
\end{align}
In the same basis, the incremental change in the Hamiltonian $dH$ is given by
\begin{align}\label{eq:HdotSupp}
    dH=\bigoplus_{k>0} dh~\partial_h H_k = \bigoplus_{k>0} dh ~\text{diag}[ 2J, 2J, 0, 0]
\end{align}
where we have used the fact that all $n$-dependence (time-dependence) comes through the parameter $h$ and $H_k(h)$ is linear in $h$.


Before making any large $L$ assumptions, we review the effects of finite chain length on the thermal state magnetisation and geodesic path. In Fig.~\ref{fig:hgeodesic}, the length of the spin chain is seen to affect the magnetisation:
\begin{align}
    \langle \sigma_z \rangle =1-\frac{2}{L}\sum_{k>0} \langle c_k^\dagger c_k \rangle + \langle c_{-k}^\dagger c_{-k} \rangle = \frac{1}{L}\sum_{k>0}\frac{E_k}{\epsilon_k}\tanh (\beta \epsilon_k).
\end{align}
For smaller $L$, the gradient in $\langle \sigma_z \rangle$ (magnetic susceptibility) is lower around the critical point, that is, the phase transition is smoothed. The optimal paths shown in the inset of Fig.~\ref{fig:hgeodesic} are established by solving the 1D geodesic equation. We find that the metric peaks around the critical point at lower temperatures and longer chain lengths, when the phase transition is most severe. Since the change in metric is proportional to the dissipation, we can predict that the optimal protocol ought to slow down around the critical point \cite{scandi2019thermodynamic}. This is indeed the case, as can be seen in the inset of Fig.~\ref{fig:hgeodesic}. When the temperature is high ($\beta = 0.01$), the thermal influences disrupt and smooth-over the phase transition, and the geodesic path approaches a linear ramping.


\begin{figure}[ht]
    \centering
    \subfloat[]{
        \includegraphics[width=0.47\linewidth]{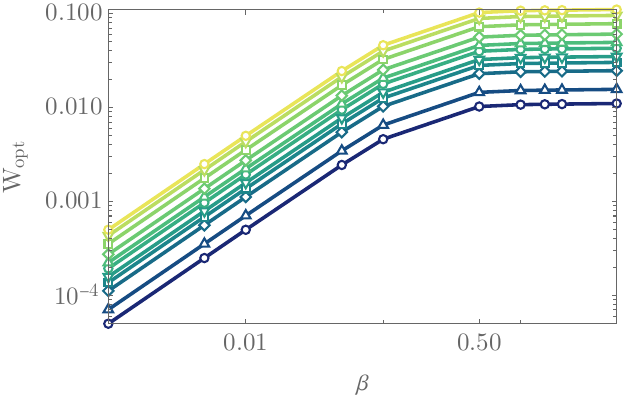}
        \label{subfig:WoptBeta}
        }
    \hfill
    \subfloat[]{
        \includegraphics[width=0.47\linewidth]{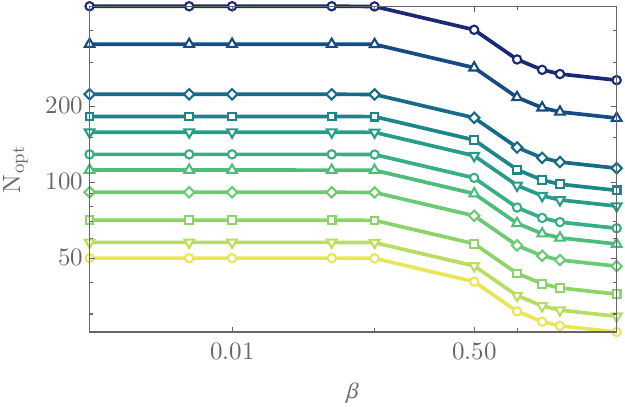}
        \label{subfig:NoptBeta}
        }
    \caption{The temperature dependence of $W_\text{opt}~[J]$ (a) and $N_\text{opt}$ (b), as described in the main text. Different colour lines refer to different noise strengths $\Phi \in [ 0.0001, 0.01]$ with $\Phi = 0.0001$ in yellow and  $\Phi = 0.01$ in dark blue.
    }
    \label{fig:OptvsBeta}
\end{figure}

Equipped with workable expression for the Hamiltonian and Gibbs state, we now endeavour to calculate the CGF for the TFIM. To streamline notation, we omit writing the explicit $h$-dependence in $\pi(h)$ and other related functions.

By recalling expressions \eqref{eq:gibbsIsingSupp} and \eqref{eq:HdotSupp}, the $y$-covariance in the definition of $K(\lambda)$, can be expressed as a single sum over $k$:
\begin{align}\label{eq:covdHdH}
     \text{cov}_{\pi}^y(dH,dH) & = (dh)^2 \sum_{k>0} C(k,y,h)
\end{align}
where
\begin{align}\label{eq:C(k,y,h)}
     C(k,y,h) = \frac{J^2 \sech (\beta \epsilon_k)}{2 \epsilon_k^2} \left[  E_k^2 + \Omega_k^2 \cosh(2\beta \epsilon_k (1-2y)) \right]
\end{align}
and $\epsilon_k =J\sqrt{h^2 + 1 -2 h \cos(k)}$.

By inserting this into the general CGF expression \eqref{eq:KdissHAM}, and looking at the dissipation per site, $\lambda/L$, by changing variable $x \rightarrow x/L$, the CGF takes the form
\begin{align}\label{eq:KdissResc2Supp}
    K\left(\frac{\lambda}{L}\right) & = -\frac{\beta^2}{2L} \sum_{n=0}^{N-1} (dh_n dh_n + \Phi_n )
    \int_0^{\lambda} dx \int_\frac{x}{L}^{1-\frac{x}{L}} dy ~\sum_{k>0} C(k,y,h_n).
\end{align}
From Eq.~\eqref{eq:KdissResc2Supp}, $W_\text{diss}/L$ (and higher order statistics) are obtained via Eq.~\eqref{eq:cumulants}---illustrated in Fig.~\ref{fig:isingWdissVsN}. If $L\gg 1$, the Riemann sum over $k$ can be replaced with an integral using $dk = 2\pi/L $, and the integral over $y$ can be split and, using the property that $C(k,y,h)$ is symmetric in $y \in [0,1]$, we get 
\begin{align}\label{eq:IsingWdissNoiseless}
    W_\text{diss} = \frac{\beta}{4 \pi} \int_0^1 \left(\frac{1}{N}\Dot{h}_t^2 + N \Phi_t \right) dt \int_0^1 dy ~ \int_0^\pi C(k,y,h_t) ~ dk.
\end{align}
where we have also moved to a continuous time description, taking $N\gg 1$. In the thermodynamic limit of $L \rightarrow \infty$, the average dissipation is thus finite. The results shown in Fig.~\ref{subfig:WoptA} are calculated from Eq.~\eqref{eq:KdissResc2Supp} and so are valid for finite $N$ and any $L$. To arrive at Eqs.~\eqref{eq:IsingWdiss} and \eqref{eq:C(k,h)}, we simply perform the integral over $y$.

Looking at Fig.~\ref{subfig:WoptA}, we see that greater $\Phi$ has increased optimal rate of dissipation, $\Dot{W}_\text{opt}$, at every time step across the erasure protocol. At intermediate and higher temperatures (Fig.~\ref{subfig:WoptB}), the geodesic path performs well in smoothing over erratic energetic losses. However, when $\beta >1$, the process is more affected by non-adiabatic transitions around the critical point ($t\approx 0.5$), causing a steep rise in dissipation rate and an overall increase in $W_\text{opt}$ by the end of the stroke (cf. Fig.~\ref{fig:isingVsPhi}). Meanwhile, when $\beta \approx 1$ or less, $\Dot{W}_\text{opt}$ is found to be largest before the critical point. The length of the spin chain is also seen to affect the magnetisation where, for smaller $L$ ($L=2$ in the inset of Fig.~\ref{subfig:WoptB}), the dissipative peak around the critical point is suppressed.

In Fig.~\ref{fig:OptvsBeta} the temperature dependence of optimal quantities is displayed, supporting claims made in the main text which we restate here:
``we see that for all noise strengths, $N_\text{opt}$ is reduced as $\beta$ increases,  changing most noticeably around $\beta \approx 1$. Therefore, it is beneficial to drive the protocol faster at lower temperatures. However, when temperatures are lowered, we can see that the average dissipated work monotonically increases, rising to a maximum value at $\beta \approx 1$. Thus, at larger temperatures, where the linear response theory applies most comfortably, $W_\text{opt}$ is minimised and $N_\text{opt}$ has the useful property of being largely insensitive to the exact temperature.''


\section{Operational construction of work statistics for noisy quenches}
\label{app:noisyTPMS}

In this appendix, we discuss the construction of work statistics equivalent to those of the two-point measurement scheme (TPMS) \cite{TalknerWorkNotObservable} with quenches subject to control noise. In the main text, our analysis is primarily concerned with the average dissipated work $W_\text{diss}$, and we do not explicitly address how this quantity might be measured. Nevertheless, such considerations are important for two reasons. First, any experimental verification of our results necessarily requires an operational measurement protocol; in quantum thermodynamics, quantities such as $W_\text{diss}$ are therefore often defined in conjunction with a specific measurement scheme. Second, while expectation values suffice for average quantities, access to higher-order work statistics—as given in the \textit{fluctuations} section and Appendix~\ref{app:CGF}—requires an explicit measurement framework.
Although continuous or weak measurement approaches could also be considered, we restrict attention here to operational schemes that reproduce the work statistics of the TPMS for the noisy driven systems studied in this work. In particular, we employ a full-counting-statistics (FCS) framework based on coherent system–meter coupling, which yields the same work distributions as the TPMS when the initial state is diagonal in the energy eigenbasis. 
Specifically, we address whether work statistics can be consistently defined with respect to the instantaneous eigenbasis of the full system Hamiltonian, including both deterministic and stochastic contributions.

In the absence of control noise, TPM and FCS constructions of quantum work statistics may be implemented without complication, since the instantaneous energy eigenbasis is exactly known. Because the initial state is thermal and hence diagonal in the energy eigenbasis, projective measurements do not alter the average energy statistics or induce additional energetic back-action on the system 
\cite{DeffnerCostMeasurement,kammerlander2016coherence}. When control noise is present, however, the instantaneous energy eigenbasis is no longer known to the observer. 
If measurements were instead performed in the noiseless eigenbasis, the system state would generally possess coherence in that basis, which would subsequently be destroyed by the measurement process, leading to spurious energetic contributions and altered dynamics. To avoid this inconsistency, it must be assumed that energy measurements are defined with respect to the instantaneous eigenbasis of the full (noisy) system Hamiltonian. Although this basis is not known a priori to the observer, this assumption admits a clear operational justification based on standard measurement constructions used in quantum work statistics.

\begin{figure}[t]
\centering
\begin{quantikz}[row sep=0.5cm, column sep=0.65cm]
\lstick{$B:\ \rho_B$}
& \qw
& \qw
& \qw
& \gate[wires=2]{U^{(n+1)}_{SB}}
& \trash{}
\\
\lstick{$S:\ \pi_n$}
& \gate[wires=2]{U^{(n)}_{SM}}
& \gate{dH_n}
& \gate[wires=2]{\big(U^{(n+1)}_{SM}\big)^\dagger}
& \qw
& \qw
\\
\lstick{$M:\ |0\rangle$}
\qw
& \qw
& \qw
& \qw
& \qw
& \meter{}
\end{quantikz}
\caption{Quantum circuit representation of a single discrete quench step \(n \to n+1\). 
The system \(S\) is initially in the Gibbs state \(\pi_n \propto e^{-\beta H_n}\) and is coherently coupled to an auxiliary meter \(M\) via a quantum non-demolition interaction \(e^{-i\lambda H_n \otimes P_M}\), encoding the pre-quench energy. 
The Hamiltonian is then suddenly updated by \(dH_n = H_{n+1}-H_n\). 
A second coupling \(e^{+i\lambda H_{n+1}\otimes P_M} \) encodes the post-quench energy into the same meter degree of freedom. 
Thermalisation is generated by a joint unitary \(U^{(n+1)}_{SB}\) acting on the system and bath \(B\); tracing out \(B\) yields relaxation toward the Gibbs state \(\pi_{n+1}\).
A final measurement of the meter yields the work statistics, with the QND couplings preserving the instantaneous energy populations and without requiring instantaneous projective measurements at the quench.}

\label{fig:quench_circuit}
\end{figure}
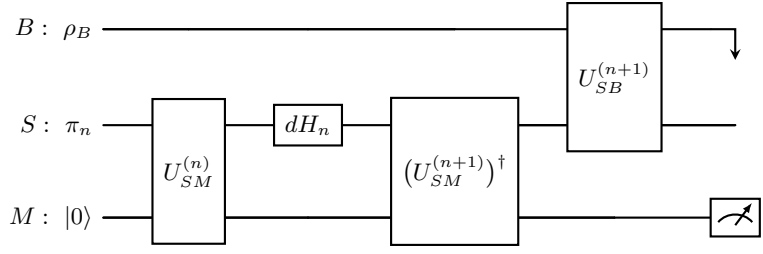

As depicted in Fig.~\ref{fig:quench_circuit}, we model energy measurements using an auxiliary quantum system (meter) that is briefly and weakly coupled to the system of interest. The system--meter interaction is taken to be of quantum non-demolition (QND) form \cite{BraginskyQND},
\begin{equation}
    H_{\mathrm{int}} = H_n \otimes P_M ,
\end{equation}
where $H_n$ now denotes the full (deterministic + stochastic) instantaneous system Hamiltonian at step $n$, and $P_M$ is a conjugate observable of the meter. Interactions like this have been observed in circuit QED scenarios (see later). 
We thus posit that the meter couples naturally to the total system Hamiltonian. 
Over a short interaction time, this generates the unitary
\begin{equation}
    U_{SM}^{(n)} = e^{- i \lambda H_n \otimes P_M} = \sum_a \Pi^a_n \otimes e^{-i \lambda E_n^a P_M},
\end{equation}
which coherently encodes the system energy into the meter without altering the system’s energy populations, since $[U_{SM}^{(n)},H_n \otimes \mathbb{I}_M]=0$. In reaching the second equality, we have used the spectral decomposition of the system Hamiltonian into its energy eigenstates $\Pi^a_n$, $H_n = \sum_a E^a_n \Pi^a_n$. Crucially, the observable being measured is fixed by the operator appearing in the coupling itself: the meter couples to the actual Hamiltonian governing the system dynamics, including stochastic fluctuations, and does not require knowledge of the Hamiltonian eigenbasis by the observer.

When the system is allowed to thermalise prior to the measurement, its state is given by the Gibbs state $\pi_n \propto e^{-\beta H_n}$, which is diagonal in the eigenbasis of $H_n$. As a result, energy measurements performed at this stage do not alter the state populations or destroy coherence and so, in the present thermodynamic accounting, do not induce additional energetic costs. This applies in particular to the pre-quench measurement in the two-point measurement protocol.

Following a sudden quench $H_n \rightarrow H_{n+1}$, the system state is generally not diagonal in the eigenbasis of $H_{n+1}$. Nevertheless, a second QND coupling of opposite sign, $\big(U_{SM}^{(n+1)}\big)^\dagger= e^{ i \lambda H_{n+1}\otimes P_M}$, may be applied immediately after the quench, without requiring relaxation or decoherence. This interaction defines an energy measurement with respect to the instantaneous Hamiltonian $H_{n+1}$ and coherently encodes the post-quench energy into the same meter degree of freedom. Importantly, no classical record of the measurement outcome is required at the time of the quench; it suffices that the system--meter coupling is well defined, with readout performed at a later stage.

Operationally, this protocol is closer to an FCS implementation of quantum work than to explicit wavefunction collapse at two separate times. Nevertheless, for thermal initial states satisfying $[\pi_n,H_n]=0$, the resulting work probability distributions coincide with those obtained from the TPMS \cite{esposito2009nonequilibrium}.
It is directly analogous to quantum non-demolition energy measurements implemented via dispersive coupling in circuit quantum electrodynamics, where a superconducting qubit is coupled to a microwave resonator whose frequency shift tracks the qubit's instantaneous (dressed) energy levels \cite{Blais2004,Gambetta2008,BianchettiQED}. In such settings, rapid parameter changes or control noise modify the effective system Hamiltonian, yet the measurement remains sensitive to the instantaneous energies generated by the actual dynamics.

The irreversibility associated with the quench arises not from the measurement itself, but from the subsequent thermalisation of the post-quench state toward the Gibbs state of $H_{n+1}$. The resulting entropy production is quantified by the relative entropy $S(\pi_n \| \pi_{n+1})$, yielding the dissipated work appearing in Eq.~\eqref{eq:relent}.

\section{Work fluctuations}
\label{app:CGF}

In this section we consider the higher order statistics of work done during the step-equilibration process. To do so we perform a series of two-point energy measurements \cite{esposito2009nonequilibrium}  across the sudden changes in controls, following the setup described in Appendix~\ref{app:noisyTPMS}
. For each step $n$, we thus construct the probability distribution ,
\begin{align}\label{eq:prob(w)}
    P_\text{diss} (w_n) & = \sum_{j,k} \delta\left[w_n-\left(E_n^k - E_n^j +\frac{1}{\beta} \log \frac{Z_{n+1}}{Z_n}\right) \right] \times\bra{E_n^j}\pi_n \ket{E_n^j} \abs{\bra{E_n^j}\ket{E_{n+1}^k}}^2,
\end{align}
where $E_n^k$ is the $k$\textsuperscript{th} eigenvalue of $H_n$ and $\ket{E_n^k}$ the corresponding eigenvector. The $\frac{1}{\beta} \log \frac{Z_{n+1}}{Z_n}$ term within the Dirac delta is the subtraction of the free energy change $\Delta F$ from the measured work, ensuring that only the dissipated work is counted.

Instead of calculating the full work probability distribution $P_\text{diss}^\text{tot}(w)$ from $N$ convolutions of all the $P_\text{diss} (w_n)$, it is more tractable to compute the cumulant generating function (CGF), which is additive for independent random variables,
\begin{align}
    K(\lambda)  := \sum_{n=0}^{N-1} \log \int_{-\infty}^\infty dw_n ~ P_\text{diss}(w_n) e^{-\beta w_n \lambda} .\label{eq:CGF2}
\end{align}
The mean and variance in dissipated work are given by deriviatives of the CGF:
\begin{align}
    &\beta W_{\text{diss}}=-\frac{d}{d\lambda}K(\lambda)\bigg|_{\lambda=0}, \\
    &\beta^2 \sigma_{W}=\frac{d^2}{d\lambda^2}K(\lambda)\bigg|_{\lambda=0},
\end{align}
After substituting in the distribution \eqref{eq:prob(w)} and simplifying, we get \cite{quantumWorkStats}
\begin{align}
    K(\lambda) & = \sum_{n=0}^{N-1} \log \Tr [ \pi_{n+1}^\lambda \pi_n^{1-\lambda} ] \label{eq:CGFgibbs}
\end{align}
Equipped with workable expressions for the Hamiltonian and linear Gibbs state expansions \eqref{eq:pi_expansion}, the CGF is approximately \cite{quantumWorkStats} 
\begin{align}\label{eq:Kdiss}
    K(\lambda) = -\frac{\beta^2}{2} \sum_{n=0}^{N-1} \int_0^\lambda dx \int_x^{1-x} dy ~ \text{cov}^y_{\pi_n} (dH_n,dH_n)
\end{align}
where the generalised $y$-covariance for non-commuting quantities is given by
\begin{align}\label{eq:covyAB}
    \text{cov}^y_\rho(A,B) = \Tr \left[ \rho^{1-y} \Delta_\rho (A)\rho^y B\right]
\end{align}
with $\Delta_\rho(A) := A - \Tr (A \rho)$. The covariance $\text{cov}^y_\rho(A,B)$ is symmetric in $A$ and $B$ and reduces to the standard covariance $\langle A B \rangle_\rho -\langle A \rangle_\rho \langle B \rangle_\rho$ when all quantities commute.

From $K(\lambda)$ in Eq.~\eqref{eq:Kdiss}, the cumulants of $W_\text{diss}$ are gathered via differentiation with respect to $\lambda$, alongside the correct $\beta$ scaling. The $j$\textsuperscript{th} cumulant is given by:
\begin{align}\label{eq:cumulants}
    \mathcal{I}_\text{diss}^{(j)} &=\left.\frac{1}{(-\beta)^j}\frac{d^j}{d \lambda^j}K(\lambda)\right|_{\lambda =0}.
\end{align}
Following the approach laid out in the main body and Appendix~\ref{app:noise}, we get an expression for the noise-averaged CGF,
\begin{align}
    K(\lambda)  & =  - \frac{\beta^2}{2}\sum_{n=0}^{N-1} \int_0^\lambda dx \int_x^{1-x} dy ~ \Big[\text{cov}^y_{\pi_n} (dH_n,dH_n)  + \myinnerp*{ \text{cov}^y_{\pi_n} (d\Tilde{H}_n,d\Tilde{H}_n ) } \Big]  +\mathcal{O}(\epsilon^3). \label{eq:KdissHAM} 
\end{align}
Again, this result follows from the substitution $dH_n \mapsto dH_n + d\Tilde{H}_n$, keeping small terms up to $\mathcal{O}(\epsilon^2)$, whereby only the 0\textsuperscript{th} order terms of $\pi_n^x$ and $\Delta_{\pi_n}(A)$ remain in Eq.~\eqref{eq:covyAB}. We can also decompose the Hamiltonian using Eq.~\eqref{eq:Ham}, giving
\begin{align}
    K(\lambda)   = - \frac{\beta^2}{2} \sum_{n=0}^{N-1} & \Big( dv_n^\alpha dv_n^\gamma  + \Phi_{n}^{\alpha \gamma}\Big) \int_0^\lambda dx  \int_x^{1-x} dy ~  \text{cov}^y_{\pi_n} (V^\alpha,V^\gamma). \label{eq:KdissDecomp} 
\end{align}
The first derivative at $\lambda=0$ recovers our result for the mean dissipated work \eqref{eq:WdissCont}. To get the variance in the linear response regime, the second derivative gives the result \eqref{eq:variance} stated in main text
\begin{align}
    \frac{\beta}{2}\myinnerp*{\sigma_W^2}  = \frac{1}{2}\int^1_0 dt \  \left( \frac{1}{N}  m_{ij}(\vec{v}_t) \Dot{v}_t^i\Dot{v}_t^j + N  m_{ij}(\vec{v}_t)\Phi^{ij}_{t} \right) 
\end{align}
where 
\begin{align}
     m_{ij}(\vec{v}_t)&=\beta \bigg(\! \frac{1}{2}\Tr\left[ \{V_i, V_j\} \pi_t \right]-\Tr\left[V_i\pi_t\right]\Tr\left[V_j\pi_t\right] \! \bigg)
\end{align}
The metric $m_{ij}$ gives rise to another form thermodynamic length \cite{MillerMehboudi2020} 
\begin{align}\label{eq:thermolengthHigher}
    \tilde{\mathcal{L}}(\Lambda) = \int_0^{1} dt \sqrt{ m_{ij}(\vec{v}_t) \Dot{v}_t^i  \Dot{v}_t^j }  
\end{align}
and hence, an optimal number of steps for minimal work fluctuations:
\begin{align}\label{eq:Noptj}
    \tilde{N}_\text{opt}= \frac{\tilde{\mathcal{L}}(\Lambda^*)}{\sqrt{ \displaystyle\int^1_0  dt \   m_{ij}(\vec{v}_t^{*}) \Phi^{ij}_{t} ~}}.
\end{align}
In general, this step number differs from Eq.~\eqref{eq:Nopt} due to the inequivalence between the metrics $g_{ij}\neq m_{ij}$. This arises from any non-commutativity across the Hamiltonian protocol, resulting in the generation of quantum friction \cite{PhysRevLett.123.230603}.

\end{document}